\documentclass[12pt]{iopart}
\usepackage{latexsym,ulem,cancel}
\usepackage{graphicx,amssymb}
\begin{document}
\title{Cross-Over between universality classes in a magnetically disordered metallic wire.}
\author{Guillaume Paulin }
\address{Institut f\"ur Theoretische Physik, Universit\"at zu K¨\"oln, Z¨\"ulpicher Str. 77, 50937 K¨\"oln, Deutschland }

\author{David Carpentier}
\address{CNRS  - Laboratoire de Physique de l'Ecole Normale
Sup{\'e}rieure de Lyon, \\
46, All{\'e}e d'Italie, 69007 Lyon, France}

%
%
\begin{abstract}
In this article we present numerical results of conduction in a disordered quasi-1D wire in the possible presence of magnetic impurities. Our analysis leads us to the study of universal properties in different conduction regimes such as the localized and metallic ones. In particular, we analyse the cross-over between universality classes occuring when the strength of magnetic disorder is increased. For this purpose, we use 
a numerical Landauer approach, and derive the scattering matrix of the wire from electron's Green's function 

 \end{abstract}

\maketitle

\section{Introduction}
 Interplay between disorder and quantum interferences leads to one of the most remarkable phenomenon in 
 condensed matter : the Anderson localization of waves. The possibility to probe directly the properties of 
 this localization with cold atoms\cite{billy:2008,roati:2008} have greatly renewed the interest on this fascinating physics.  
  In this paper, we focus on the particular situation where electrons encounter two kinds of disorder:  a usual scalar potential at the origin of diffusion, and a magnetic potential, arising from a collection of frozen random magnetic moments. This situation is naturally realized experimentally in the study of transport properties 
  of metallic spin glass wires 
 \cite{deVegvar:1991,Jaroszynski:1998,Neuttiens:1998,Capron:2011}. In these wires, the spins freeze at low temperatures when entering the spin glass phase due to the frustrating magnetic couplings. In this glassy phase, and neglecting any residual Kondo effect in this regime, the impurities act effectively as a (weak) magnetic potential. We study numerically the effect of both types of disorder on the statistical properties of the wire conductance. In particular, we will focus on the experimentally relevant crossover of (weak) localization properties of the wire as a function of the magnetic disorder strength.

  One dimensional disordered electronic systems are always localized. Following the scaling theory
  \cite{Abrahams:1979} this implies that by increasing the length $L_{x}$ of the wire for a fixed 
  amplitude of disorder, its typical conductance ultimately reaches  vanishingly small values. The 
  localization length $\xi$ separates metallic regime for small length $L_{x}\ll \xi $ from the 
  asymptotic insulating regime. In the present paper, we focus on several universal properties of 
  both metallic and insulating regime of these wires in the simultaneous presence of two kinds of 
  disorder. 
    The first type corresponds to scalar potentials induced by the impurities, for which the system 
 has time reversal symmetry (TRS) and spin rotation degeneracy. In this class the Hamiltonian 
 belongs to the so-called  Gaussian Orthogonal Ensemble (GOE) of the Random Matrix Theory 
 classification \cite{Evers:2008} (RMT), corresponding to the class AI in the modern classification of Anderson universality classes
  (see {\it e.g.} \cite{Ryu:2010}). 
 If impurities do have a spin, the TRS is broken as well as spin rotation invariance. The Hamiltonian
  is then a unitary matrix, which corresponds in RMT to the Gaussian Unitary Ensemble (GUE) with 
  the breaking of Kramers degeneracy \cite{Mirlin:2000}, and to the Anderson class A \cite{Ryu:2010}. 
  However, for the experimentally relevant 
  case of a magnetic potential  weaker than the scalar potential, the system is neither described by the 
  GUE class, nor by the GOE class, but extrapolates in between. This intermediate regime, of particular relevance experimentally, 
  is the main object of study of the present paper. Moreover the present work 
  paved the way towards a numerical study of the correlation of conductances in the cross over regime \cite{Paulin:2011b}. 
  \\
   This paper is organized as follows : 
  in section 2, the model and the numerical method used will be described in details. In section 3 we identify the localized and metallic regime of transport of the system. The 
  localization length will be determined by two different methods and the cross over between both 
  universality classes (GOE, GUE) will be highlighted. In section 4, the insulating regime is studied with a particular focus on the statistical distribution of the conductance, which
allows us to highlight universal behavior. In section 5, we turn to the study of the metallic regime, and perform a careful analysis of each of the
first three cumulants of the statistical distribution of the
conductance. We focus on the universal properties of conductance
fluctuations, and the non-analyticity  of the complete distribution is
discussed. Finally section 6 is devoted to the conclusion.
  
\section{The model and the method}
\subsection{The model}

In this paper, we study numerically the scaling of transport properties of wires in the presence of magnetic and scalar disorders. We will focus on the regime of phase coherent transport, reached experimentally at low temperature (see in particular \cite{Capron:2011}). In this regime, the phase coherence length $L_{\phi}$ which phenomenologically 
accounts for inelastic scattering of electrons on impurities \cite{Akkermans:2007} is larger than (or comparable with) the wire's length $L_x$, so that 
phase coherence for the propagating electrons is preserved in the whole sample. Note that this phase coherence length $L_{\phi}$ includes in particular a contribution from inelastic scattering on the non frozen magnetic impurities through a Kondo dephasing, which is strongly reduced when entering the magnetic glass phase \cite{Capron:2011}. 

 We describe the behavior of electrons inside the disordered wire using a tight-binding Anderson lattice model with 
two kinds of disorder potentials:  
\begin{eqnarray}
\label{eq:Ham}
\mathcal{H} = \sum_{<i,j>,s}t_{ij}c_{j,s}^{\dagger}c_{i,s} &+& \sum_{i,s}v_ic_{i,s}^{\dagger}c_{i,s} 
+ J\sum_{i,s,s'}{\vec{S}}_i .  {\vec{\sigma}}_{s,s'}c_{i,s}^{\dagger}c_{i,s'}.
\end{eqnarray}
$t_{ij}$ is the hopping term from site $i$ to $j$. In the following, $t_{ij}$ will take two different values: 
$t_{ij} = t_{\slash\slash}$  in the longitudinal $x$ direction and $t_{ij} = t_{\perp}$ 
 in the transverse $y$ direction. The  scalar disorder potential $V=\{v_i\}_i$ is diagonal 
in electron-spin space. We choose the $v_i$ to be random scalars uniformly distributed in the interval $[-W/2,W/2]$. In this work, we have chosen without loss of generality to fix $t_{\slash\slash}=1$ so that all energy scales are relative to the bandwidth $t_{\slash\slash}=1$, and the amplitude of disorder $W = 0.6$. 
In eq.~(\ref{eq:Ham}) 
$s,s'$ label the $SU(2)$ spin of electrons and the $\vec{S}_i$ account for spins of the frozen magnetic impurities. A realistic choice for these frozen spins in a spin glass phase consist in considering 
classical spins with random orientations, thereby neglecting any small spatial correlation in a spin glass configuration \cite{Mezard:1987}. 
 The coupling $J$ between the electron spins and the magnetic impurities fixes the amplitude of the magnetic disorder. In this work, we will monitor the behavior of the transport properties of the sample as a function of this amplitude $J$, from $J=0$ to $J=0.4$. Indeed, 
variation of the amplitude of magnetic disorder $J$ allows to extrapolate from GOE / class AI for $J=0$ to GUE / class A for $J\neq 0$. 
 We will also show in section 5 that, as a bonus, the presence of the magnetic disorder allows also an numerically easier settlement of the universal metallic regime of the wire.

 For a given realization of both scalar $\{v_i\}_i$ and magnetic disorder $\{\vec{S}_i\}_i$, 
the Landauer conductance of this model on a 2D square lattice of size $L_{x}\times L_{y}$ is evaluated numerically using a recursive Landaueur  method  described in details in the next paragraph.

\subsection{The method}

 We have chosen to use a numerical method based on the lattice model (\ref{eq:Ham}) as opposed to e.g. random matrix or 
 Dorokhov-Mello-Pereyra-Kumar (DMPK) \cite{Dorokhov:1983,Mello:1988} method so as to provide a numerical study allowing for an easy comparison with the experimental situation\cite{Capron:2011}. Moreover, this method allows for further natural developments such as the  study of the conductance change upon magnetic impurities spin flipping, which would be difficult to reach by alternative method. Starting from a lattice model such as (\ref{eq:Ham}), the natural method providing the conductance of a finite size sample is based on the Landauer formalism \cite{Landauer:1957}. 

We consider a two-terminal setup, with electrodes connected to
the wire at $x=0$ and at $x=L_x$. These electrodes are described as semi-infinite ribbons with the same transverse geometry as the sample, and described with (\ref{eq:Ham}) but without randomness. 
 Electrons are then confined in the transverse $y$-direction via a potential that has the form:
\begin{eqnarray}
V(y) &=& 0\hspace{0.5cm}\mathrm{if}\hspace{0.5cm} 0\leq y \leq
L_y
\ ; \ 
V(y) =\infty\hspace{0.5cm}\mathrm{otherwise}
\end{eqnarray}
This confining potential in the $y$ direction leads to the appearance in the electrodes of $N$  modes propagating in the $x$-direction. 
The complete wave function of an electron in this tight binding lattice model reads then:
\begin{equation}
\label{equ:eigen_fct}
\psi(x,y) = \phi_n(y)e^{\imath k_xx},
\end{equation}
where $k_x$ is the momentum of electrons in the longitudinal direction and 
\begin{equation}
\phi_n(y) = \sqrt{\frac{2}{N_y+1}}\sin\left(\frac{n\pi y}{N_y+1}\right).
\end{equation}
We used $N_y = L_y$ in units of lattice spacing. The group velocity of this mode reads:
\begin{equation}
v_n = 2 \frac{t_{\slash\slash}}{\hbar}\sin\left(k_x\right).
\end{equation}
where $t_{\slash\slash}$ is the longitudinal hopping amplitude. This velocity depends on the momentum 
$k_{x}$ which is determined for a constant energy by the dispersion relation 
\begin{equation}
E_n = \mu - 2t_{\perp}\cos\left(\frac{n\pi}{N_y + 1}\right) -2t_{\slash\slash}\cos(k_x).
\end{equation}
At given energy $E - \mu$, we end up with the following relation for the longitudinal part of the momentum of the electron:
\begin{equation}
k_x = \arccos\left(\frac{\mu-E}{2t_{\slash\slash}}-\frac{t_{\perp}}{t_{\slash\slash}}\cos\left(\frac{n\pi}{N_y + 1}\right)\right).
\label{equ:k_x}
\end{equation}
  To optimize the efficiency of the numerical study, we fix $t_{\slash\slash}= 2 t_{\perp}$ and stay near the band center, avoiding in particular the presence of fluctuating states studied in \cite{Deych:2003}.

To compute the conductance of such a wire, the Landauer-B\"uttiker formalism of coherent transport 
is used \cite{Buttiker:1985}. It allows to relate the dimensionless conductance $g$ of a diffusive wire with the scattering matrix $T$:
\begin{equation}
g = \sum_{\mathrm{modes}\,m,n} T_{mn},
\end{equation}
where $T_{mn}$ is the transmission coefficient between modes $m$ and $n$. 
The dimensionless conductance $g$ is defined from the conductance $G$ of the system as 
$g=G / (2 e^2/h)$ when spin degeneracy is present ($J=0$), and $g=G / ( e^2/h)$ otherwise ($J\neq 0$). 

\begin{figure}[ht]
\centerline{
\includegraphics[width=13cm]{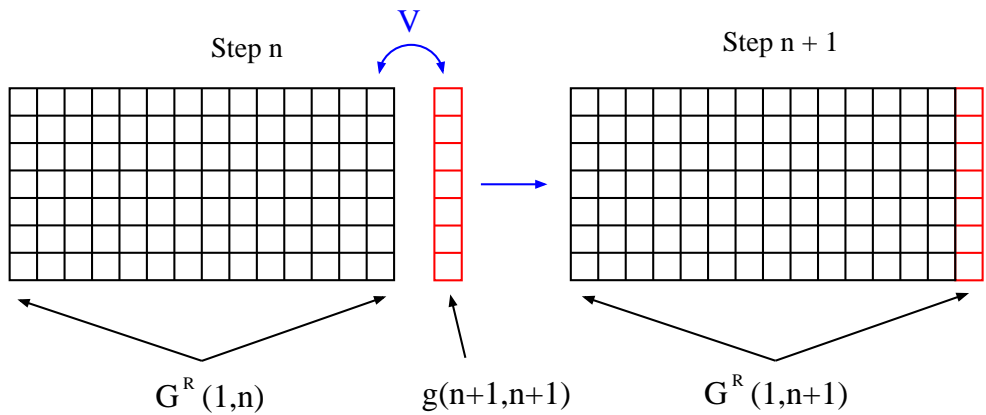}}
\caption{\label{fig:recursive}
Principle of recursive calculation of retarded Green's functions of the wire. Use of a Dyson:  $G^R_{n+1} =G^R_n + G^R_nVG^R_{n+1}$. 
At each step the longitudinal length $L_x$ is increased by one lattice spacing. 
}
\end{figure}

\begin{figure}[ht]
\centerline{
\includegraphics[width=13cm]{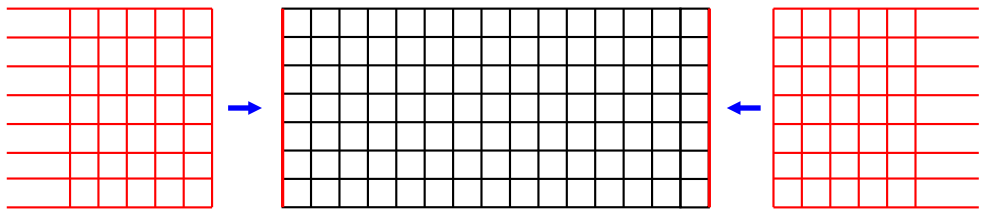}}
\caption{\label{fig:boundaries}
Boundary conditions:  the wire is connected to two leads represented by two semi infinite metallic wires.
}
\end{figure}

Following Fisher and Lee\cite{Fisher:1981} we relate the scattering matrix to the electronic retarded 
Green's functions of the system through:  
 \begin{eqnarray}
 t_{mn} &=&  \imath\hbar\sqrt{v_nv_m}\sum_{y,y'=1}^{N_y}\phi_n(y)
\mathcal{G}^R(y,x = 0|y',x = L_x)\phi_m(y').
 \label{equ:FL}
 \end{eqnarray}
with $T = Tr(t^{\dagger}t)$, $v_n$ and $\phi_n$ (resp. $v_m$ and $\phi_m$) are the group velocity and the eigen wave function 
of propagating mode $n$ (resp. $m$). Mode $n$ belongs to the left lead whereas  mode $m$ belongs to the right one. 
In (\ref{equ:FL}) $\mathcal{G}^R(y,x = 0|y',x = L_x)$ represents the retarded Green's function of an electron 
between the point $(x=0,y)$ in the left contact between the system and the electrode and the point $(x=L_x,y')$ in the right contact. 
Consequently, by calculating the retarded Green's functions of the system only between its both sides, 
we are able to determine the dimensionless conductance of the wire. An efficient method 
to calculate $\mathcal{G}^R(y,x = 0|y',x = L_x)$, which takes advantage of the quasi-one dimensional nature of the ribbon consists in obtaining it recursively\cite{Croy:2006}. 
 In figure \ref{fig:recursive} the principle of the method is sketched:  using a Dyson equation we deduce 
 the boundary Green's function of a system of size $n+1$ from the corresponding Green's function of a subsystem of size $n$, and the exact Green's function of the $n+1$ row. This allows to perform matrix inversion only of the simple row system. At both initial and final steps, we reconnect the system to the semi-infinite electrodes (see fig.~\ref{fig:boundaries}) described by a standard self energy:  
\begin{equation}
\mathcal{G}^R_{\mathrm{bound}}(y_1,y_2) = -\frac{1}{t_{\slash\slash}}\sum_{n=1}^{N_y}\phi_n(y_1)e^{\imath k_x} \phi_n(y_2).
\label{equ:Green_bords}
\end{equation}
 The last step consists in combining the Green's functions of the sample of desired longitudinal size with the one of the right lead, 
 which is also given by equation (\ref{equ:Green_bords}). 
The unitarity of the corresponding scattering matrix $\{T_{mn}\}_{m,n}$ is used to monitor the accuracy of the numerical method. Such test yielded typical relative error of order $10^{-4}$ for a system 
size $L_x=1600$ and $L_y=40$. 
%
This method allows one to compute the conductance of a wire of length $L_x$ and of width $L_y$ 
for any given configuration of scalar disorder $V$ and for any configuration of frozen classical spins $\{\vec{S}_i\}_i$. 

In the next sections, we study the properties of this conductance for one given configuration of 
magnetic disorder but for many different configurations of scalar disorder. Universal properties are identified 
by varying the transverse length $L_y$ from $10$ to $80$, with the aspect ratio $L_x/L_y$ taken from $1$ to $6000$. 
Typical number of configurations of scalar disorder $V$  we used were $N_{d}=5000$, with 
exceptions for the study of the localization properties where for $L_y=10$ we sampled the conductance distribution for  $50 000$ different configurations of disorder. 

\section{Localization length}
\subsection{Determination of $\xi$}

 We start our analysis by a determination of the parameters corresponding to the metallic and localized regimes, through a careful determination of the localization length of the system. While experimentally the only accessible regime of a phase coherent wire is the metallic regime, numerically this regime is difficult to reach and describe quantitatively, as opposed to the localized regime. This is related to the extreme reduction in the number of propagating modes in the numerical system which is associated with a corresponding reduction of the localization length. Hence in order  to clearly identify the conditions to access universal properties of the transport in the metallic regime, we start by a detailed determination of this localization length in the experimentally relevant crossover situation. Afterwards we will take the opportunity of the present study to describe other characteristics of the localized regimes in the crossover situation, before turning to our main interest : the universal metallic regime.

The localization length separates short wires of metallic behavior from  
 a insulating long wires. 
A first method to access to the localization length $\xi$ from the conductance consists in considering the scaling behavior of the typical conductance $g_{typ}$ defined as:
\begin{equation}
g_{typ} = e^{\langle \log g \rangle},
\end{equation}
where $\langle\cdot\rangle$ represents the 
average over the different configurations of scalar disorder $V$.
This typical conductance decays exponentially with the longitudinal length of the wire\cite{Beenakker:1997,Slevin:2001} :  
\begin{equation}
g_{typ} \sim e^{-\frac{2L_x}{\xi}}
\end{equation}
in the regime of long wires $L_x\gg \xi$.
\begin{figure}[ht]
\centerline{\includegraphics[width=13cm]{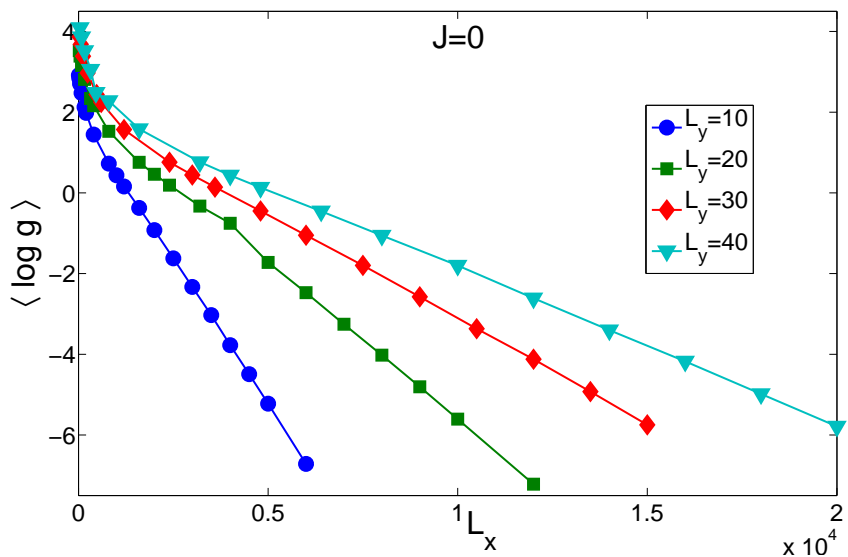}}
\caption{\label{fig:det_xi_J0}
Evolution of $\langle \log g \rangle$ as a function of longitudinal size for different transverse length ($L_y=10, 20, 30, 40$) and $J=0$. The linear part of the curve allows one to get the localization length $\xi(J=0)$ from the scaling form in the insulating regime $\langle \log g \rangle = -\frac{2L_x}{\xi}$.
}
\end{figure}
\begin{figure}[ht]
\centerline{\includegraphics[width=13cm]{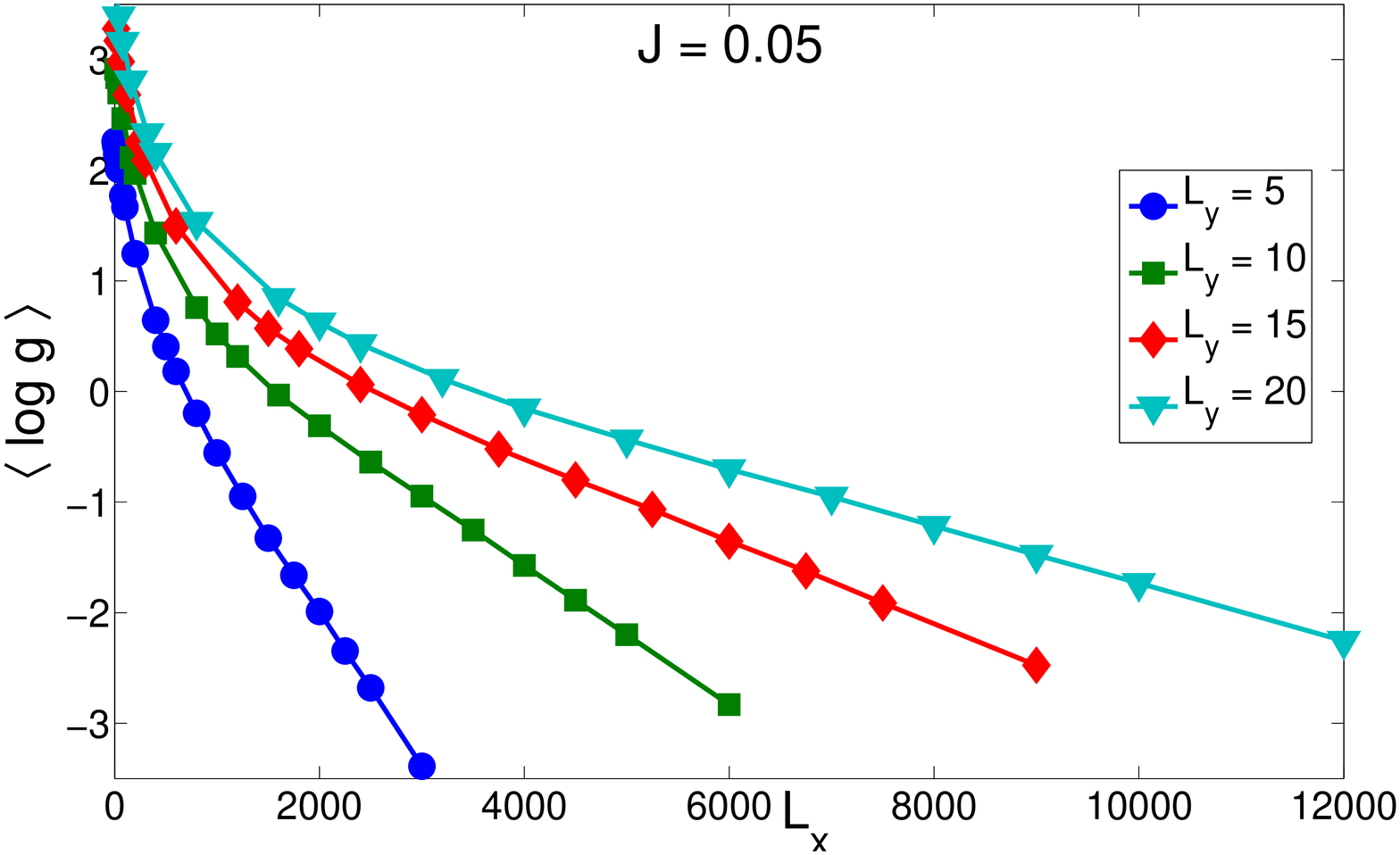}}
\caption{\label{fig:det_xi_J0p05}
Evolution of $\langle \log g \rangle$ as a function of longitudinal size for different transverse length ($L_y=10, 20, 30, 40$) and $J=0.05$. The linear part of the curve allows one to get the localization length $\xi(J=0.05)$ from the scaling form in the insulating regime $\langle \log g \rangle = -\frac{2L_x}{\xi}$.
}
\end{figure}
\begin{figure}[ht]
\centerline{\includegraphics[width=13cm]{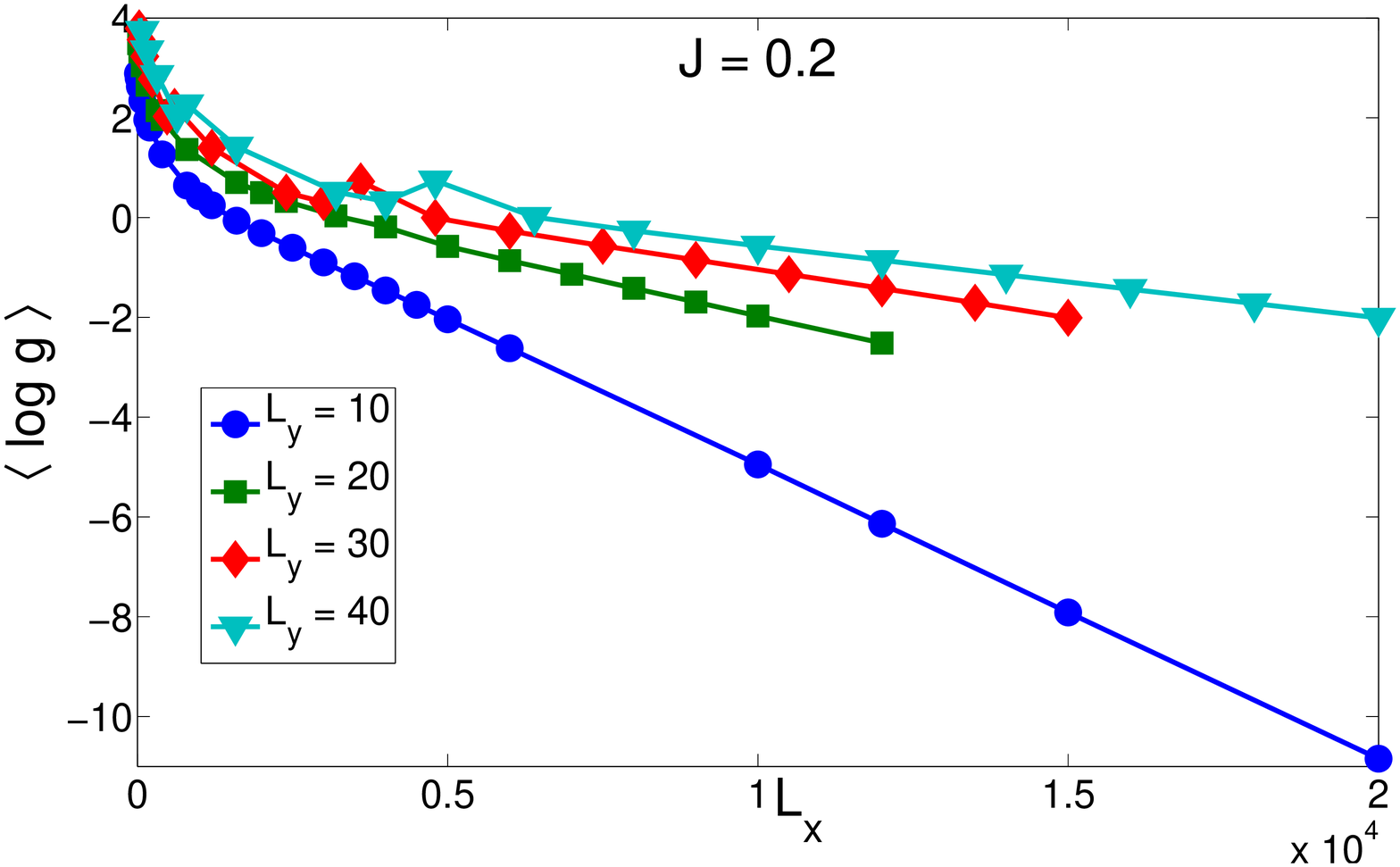}}
\caption{\label{fig:det_xi_J0p2}
Evolution of $\langle \log g \rangle$ as a function of longitudinal size for different transverse length ($L_y=10, 20, 30, 40$) and $J=0.2$. The linear part of the curve allows one to get the localization length $\xi(J=0.2)$ from the scaling form in the insulating regime $\langle \log g \rangle = -\frac{2L_x}{\xi}$.
}
\end{figure}
\begin{figure}[ht]
\centerline{\includegraphics[width=13cm]{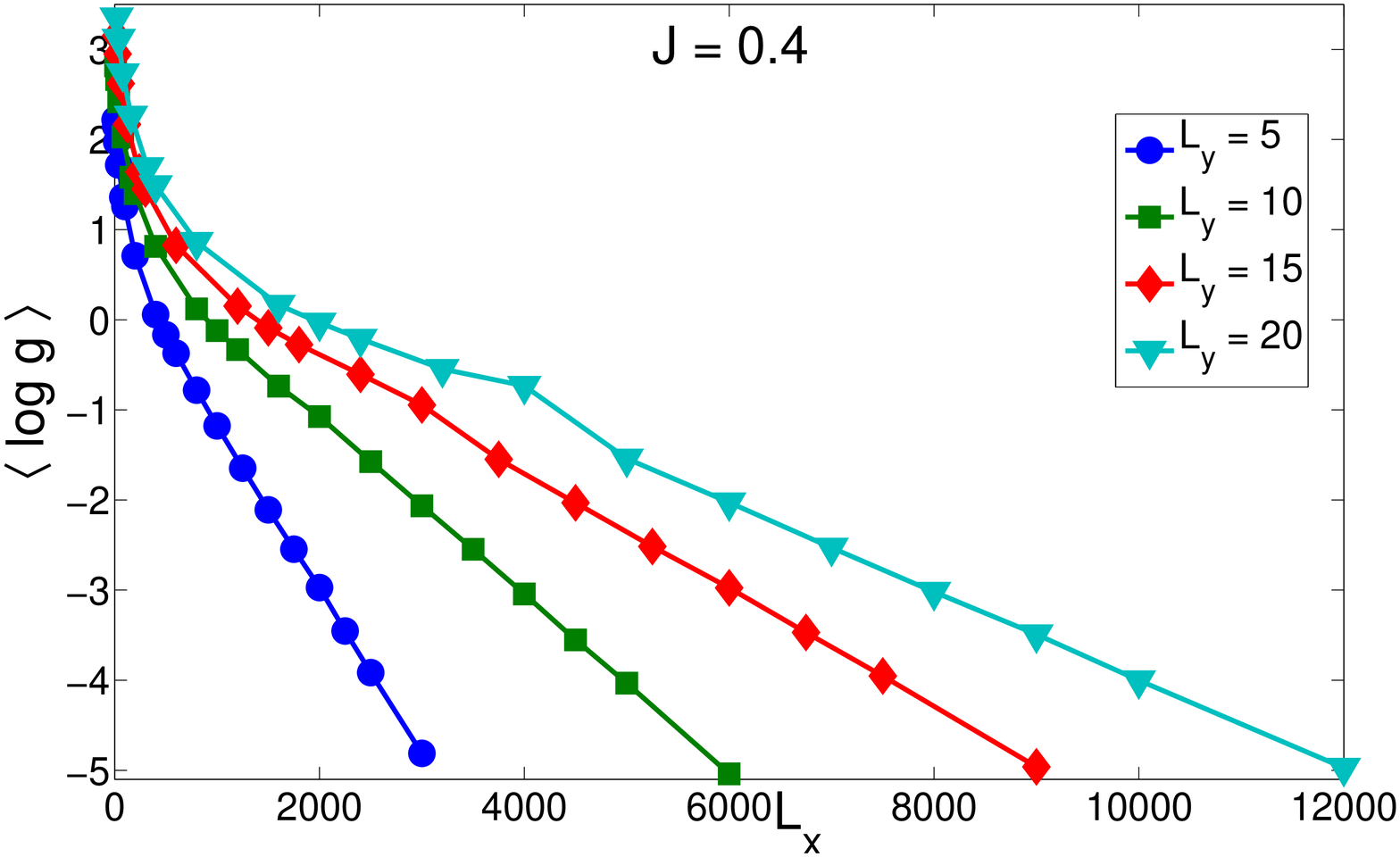}}
\caption{\label{fig:det_xi_J0p4}
Evolution of $\langle \log g \rangle$ as a function of longitudinal size for different transverse length ($L_y=10, 20, 30, 40$) and $J=0.4$. The linear part of the curve allows one to get the localization length $\xi(J=0.4)$ from the scaling form in the insulating regime $\langle \log g \rangle = -\frac{2L_x}{\xi}$.
}
\end{figure}
Figures \ref{fig:det_xi_J0}, \ref{fig:det_xi_J0p05}, \ref{fig:det_xi_J0p2} and \ref{fig:det_xi_J0p4} show the behavior of $\langle\log g\rangle$ as a function of longitudinal length $L_{x}$ for different widths $L_{y}$. Different curves correspond to different values of magnetic disorder. The linear fit of the large length part of the curves allow for a precise determination of the corresponding localization length 
for each value of $L_y$ and $J$. \\
A second method to determine this localization length consists in considering the Lyapunov exponent $\gamma$ of the transfer matrix of the system, following a standard random matrix theory approach\cite{Markos:1993,Beenakker:1997,Pichard:2001}. This exponent can be deduced from the conductance as   
\begin{equation}
\gamma(L_x) = \frac{1}{2L_x}\log\left(1 + \frac{1}{g(L_x)}\right), 
\end{equation}
 and the localization length follows from its asymptotic behavior : 
\begin{equation}
\xi^{-1} = \lim_{L_{x}\to \infty}\gamma(L_x) .
\end{equation}
On figure \ref{fig:Lyapunov_Ly10} and \ref{fig:Lyapunov_Ly20} we have plotted the Lyapunov exponent versus the inverse of the longitudinal length for different values of magnetic disorder. Different curves correspond to different widths of the wire.
\begin{figure}[ht]
\centerline{\includegraphics[width=13cm]{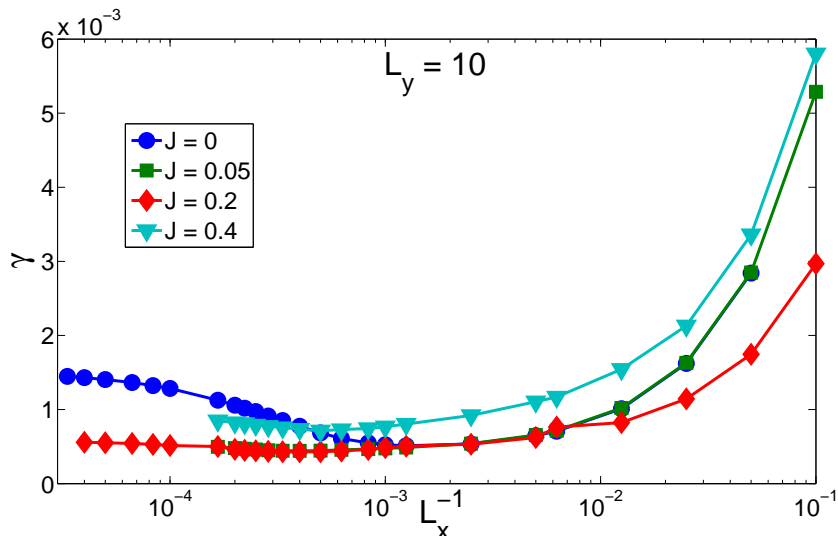}}
\caption{\label{fig:Lyapunov_Ly10}
Evolution of Lyapunov exponent with the inverse of longitudinal length in semi-log plot. Circles correspond to $J=0$, squares to $J=0.05$, diamonds to $J=0.2$ and triangles to $J=0.4$. The value of the transverse length is $L_y=10$. The localization length can be extrapolated from the value of $\gamma$ for $L_x\to\infty$.
}
\end{figure}
\begin{figure}[ht]
\centerline{\includegraphics[width=13cm]{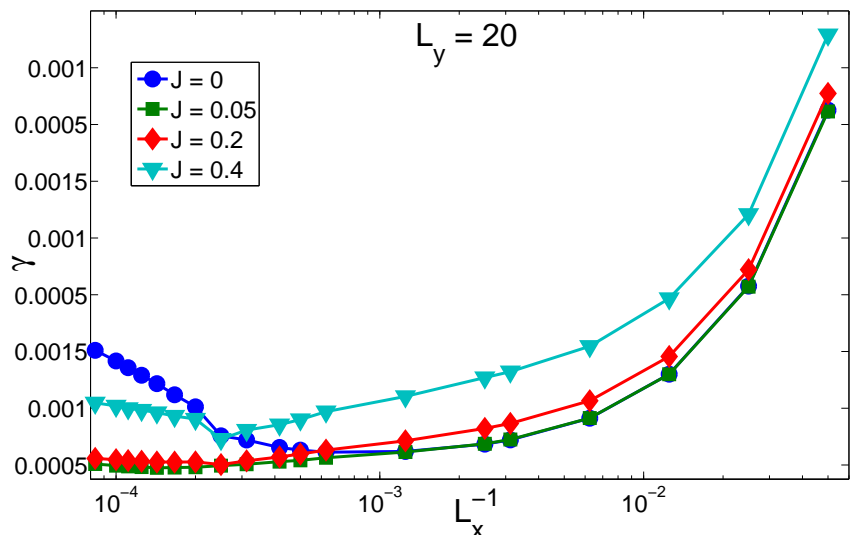}}
\caption{\label{fig:Lyapunov_Ly20}
Evolution of Lyapunov exponent with the inverse of longitudinal length in semi-log plot. Circles correspond to $J=0$, squares to $J=0.05$, diamonds to $J=0.2$ and triangles to $J=0.4$. The value of the transverse length is $L_y=20$. The localization length can be extrapolated from the value of $\gamma$ for $L_x\to\infty$.
}
\end{figure}
With this method, a simple extrapolation of the curve is necessary to obtain $\xi$, without any fit. 
the value of the inverse of the localization length for an infinite wire. Nevertheless, we have found that 
this method shows less accuracy than the preceding one: as shown on figures \ref{fig:Lyapunov_Ly10} or \ref{fig:Lyapunov_Ly20}, the Lyapunov exponent is still varying for the longest longitudinal length. We find 
that both methods give fully compatible results while the Lyapunov exponent method requires much larger system sizes than the typical conductance method for a given required accuracy. 

\begin{figure}[ht]
\centerline{\includegraphics[width=13cm]{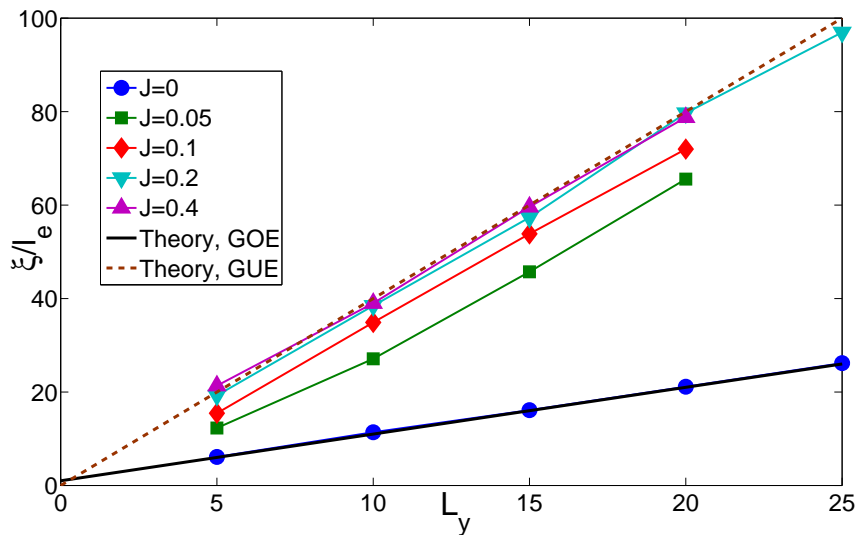}}
\caption{\label{fig:univ_class}
Evolution of localization length as a function of transverse length. $l_e$ is the mean free path of the diffusive sample. Different behavior of the localization length if $J=0$ or $J \neq 0$.
Inset : Scaling of the typical conductance $\langle \log g \rangle = -\frac{2L_x}{\xi}$.
}
\end{figure}

A first manifestation of the universality of the Anderson localization classes appears through the 
dependance of $\xi$ on the transverse length $L_{y}$ (or the number of propagating modes). 
It is expected to follow\cite{Beenakker:1997,Pichard:1991}:
\begin{equation}
\xi = (\beta L_y + 2 - \beta)l_e, 
\label{equ:xi_gen}
\end{equation}
 with $l_{e}$ the elastic mean free path and $\beta$ encodes the universal class of the model : 
 $\beta=1$ corresponds to the orthogonal universality class GOE
 while $\beta=2$ for GUE. Note that this change in $\beta$ is accompanied by an artificial 
 doubling of the number of transverse modes $N_y\equiv L_y\to 2N_y$ due to the breaking of Kramers degeneracy \cite{Beenakker:1997}. 
This effective factor 4 when breaking the spin rotation symmetry has been 
discussed in ref.~ \cite{Larkin:1983,Lerner:1995}  in details, when discussing the magnetic field 
dependance of this localization length, in comparison with random matrix 
and Non-linear sigma models. 
 Comparison of numerical localization lengths for different $J$ with (\ref{equ:xi_gen}) is 
 shown in fig~\ref{fig:univ_class}. 
 Excellent agreement is found for $J=0$ (GOE class, $\beta=1$) and with the GUE class for $J\geq 0.2$. 
 For intermediate values of $J \neq 0$ we observe a crossover between the two extreme GOE and GUE laws, which cannot be described by eq.~(\ref{equ:xi_gen}), and for which no analytical work exists to our knowledge. 
  
  From these results, we also notice that the localization regime is reached for much longer 
 wires in the presence of magnetic impurities  (GUE case)  than without (GOE):  localization is hampered by the presence of these magnetic impurities.
  As we will discuss below, this property helps in observing numerically the universal weak localization regime and the associated universal conductance fluctuations.   

\subsection{The Insulating and Metallic regimes}

\begin{figure}[ht]
\centerline{\includegraphics[width=13cm]{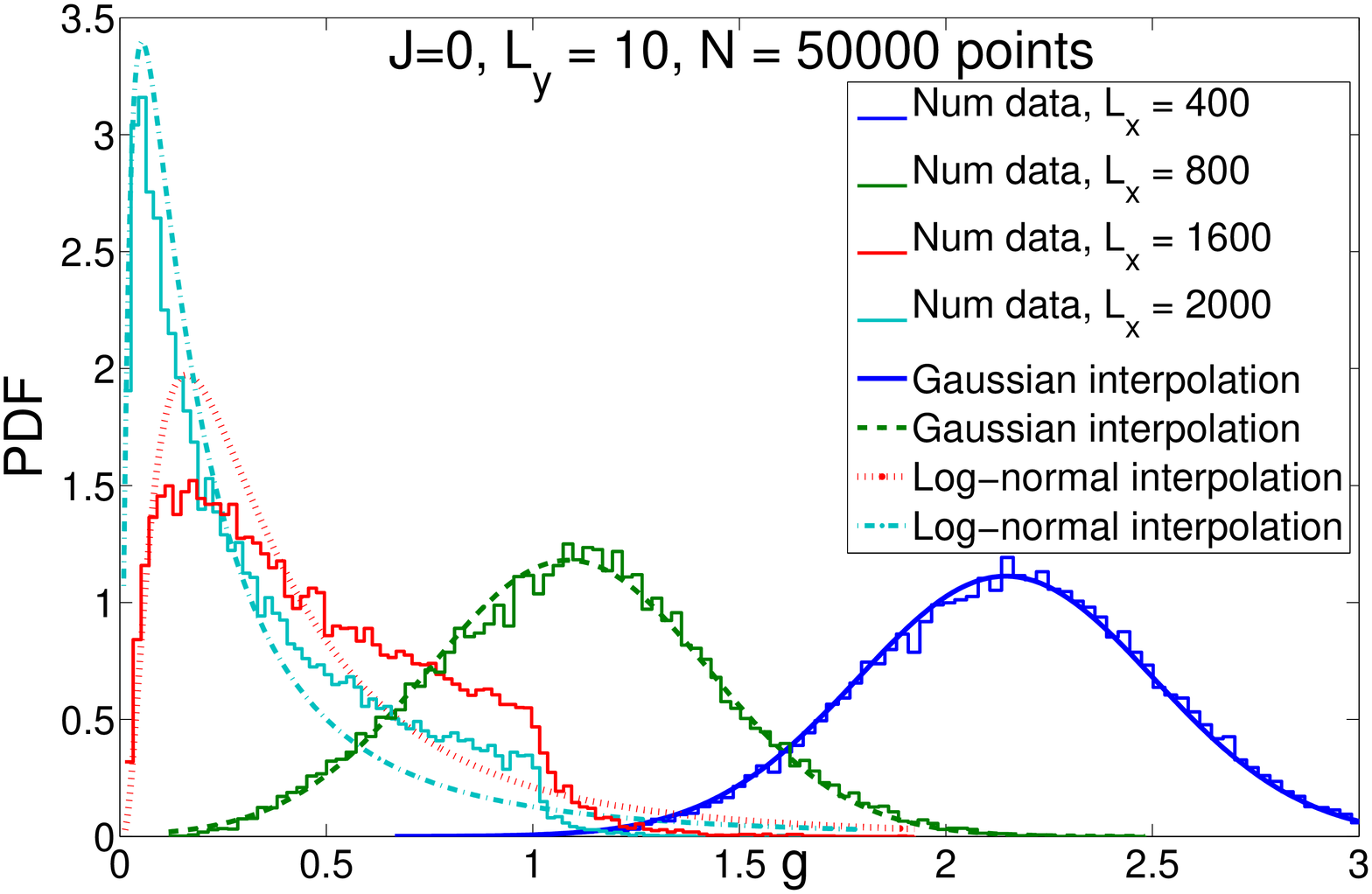}}
\caption{\label{fig:evol_PDF}
Evolution of the statistical distribution of the conductance for different longitudinal sizes for $L_y=10$ and $J=0$. $N$ is the number of disorder configurations used. Plain lines are gaussian 
(if $\langle g \rangle>1$) or log-normal (if $\langle g \rangle<1$) interpolations of numerical data.
}
\end{figure}

The localization length discriminates between both insulating and metallic regimes:  the ribbon behaves indeed as a metal ($g\gg 1$) for lengths $L_x\ll\xi$ and as an insulator ($g\ll 1$) if $L_x\gg\xi$. 
 In both asymptotic regimes the shape 
of the Probability Density Function (PDF) of the conductance which is known:  it is 
Log-normal for insulating wires\cite{Beenakker:1997} and gaussian for metallic ones.  
 By varying the longitudinal length we can study the evolution of this PDF from a gaussian 
to a log-normal distribution, as shown on figure \ref{fig:evol_PDF}. This plot is done for a given value of $L_y$ and $J$. 
One can notice that in the metallic regime, the PDF is very well approximated by a gaussian for relatively small wires : the gaussian regimes is easily reached, 
 whereas it takes length much larger than the localization length for the distribution to become 
log-normal in the insulating regime. 
This point will be discussed more precisely below on the cumulants of this PDF. 
The insulating regime is then characterized by $\langle g \rangle<1$ and the metallic one by $\langle g \rangle>1$. 

\begin{figure}[ht]
\centerline{\includegraphics[width=13cm]{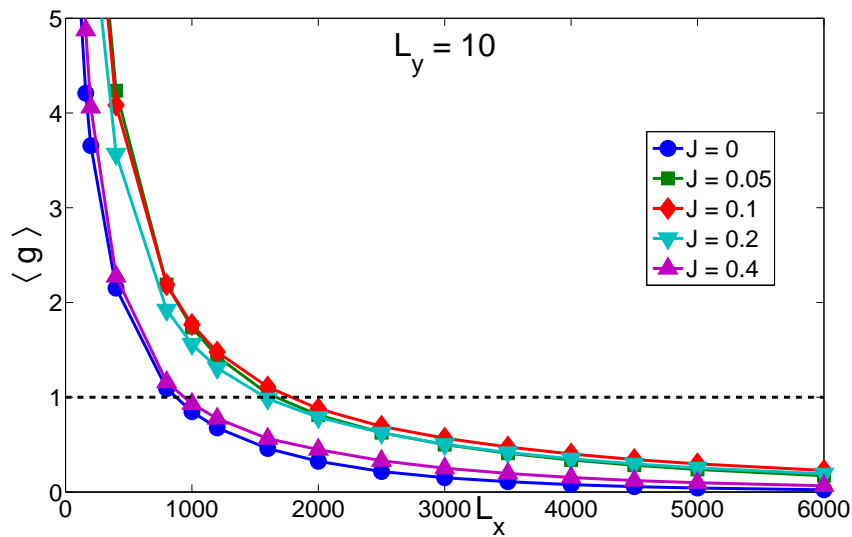}}
\caption{\label{fig:gmoyen_L}
Evolution of average conductance versus longitudinal length for different values of magnetic disorder for $L_y=10$ and $J=0,0.05, 0.1, 0.2$ and $0.4$. The line $\langle g \rangle = 1$ is plotted as a frontier between insulating and metallic regimes.
}
\end{figure}

\begin{figure}[ht]
\centerline{\includegraphics[width=13cm]{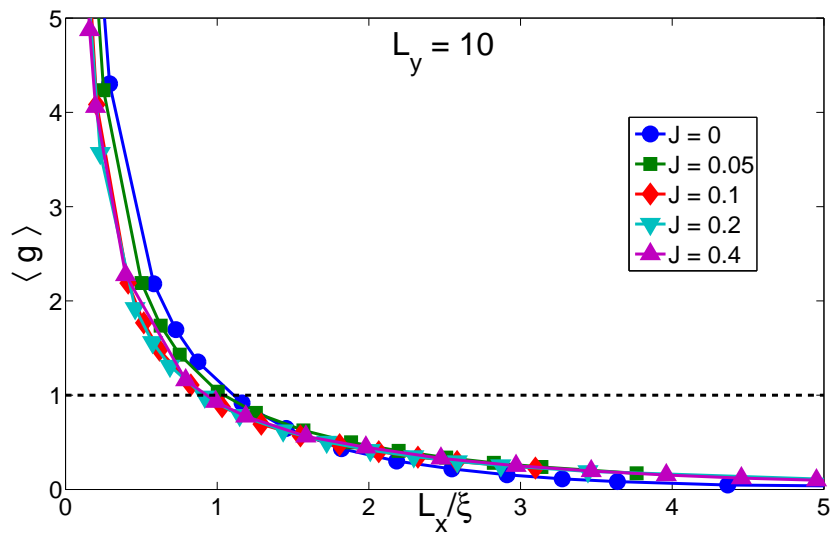}}
\caption{\label{fig:gmoyen_Loverxi}
Evolution of average conductance versus the scaling variable $L_x/\xi$, for $L_y=10$ and $J=0,0.05, 0.1, 0.2$ and $0.4$. All curves nearly collapse in one single curve. The frontier between insulating and metallic is drawn again.
}
\end{figure}

 In order to characterize samples by the average $\langle g \rangle$, and in particular plots higher cumulants as a function of $\langle g \rangle$, we now turn to a short study of the behavior of this first cumulant as a function of the system size. 
On figure \ref{fig:gmoyen_L} we have plotted $\langle g \rangle (L_{x})$  for 
different values of magnetic disorder $J$ (hence different localization lengths). 
These curves approximately collapse when plotted against the scaling variable $L_x/\xi$,  as show on fig. \ref{fig:gmoyen_Loverxi}. we remind the reader that for
$J = 0$, the average conductance is defined by $g = G/2G_0$, which explains why $J=0$ and $J=0.4$ curves coincide on  
fig.~\ref{fig:gmoyen_L}:  according to figure \ref{fig:univ_class}, only for these values of magnetic disorder universality classes are 
reached. 

This study of $\langle g \rangle (L_{x})$ allows to proceed in the study of higher cumulants of the PDF of $g$ and test prediction of the single parameter scaling of distributions\cite{MacKinnon:1981}.

\section{Universal Insulating Regime}
\subsection{Probability Density Functions}

 In the insulating regime $L_{x}\geq \xi $, we expect a Log-normal conductance statistical distribution, as seen previously.
  However in the weakly insulating regime  $\langle g \rangle\lesssim 1$ this log-normal asymptotic form is not reached. Instead, as shown in fig.~\ref{fig:evol_PDF} we find a non-analytical behavior of $P(g)$ in agreement with \cite{Muttalib:1999,Markos:2002,Froufe:2002,Muttalib:2003}. 
\begin{figure}[ht!]
\centerline{\includegraphics[width=13cm]{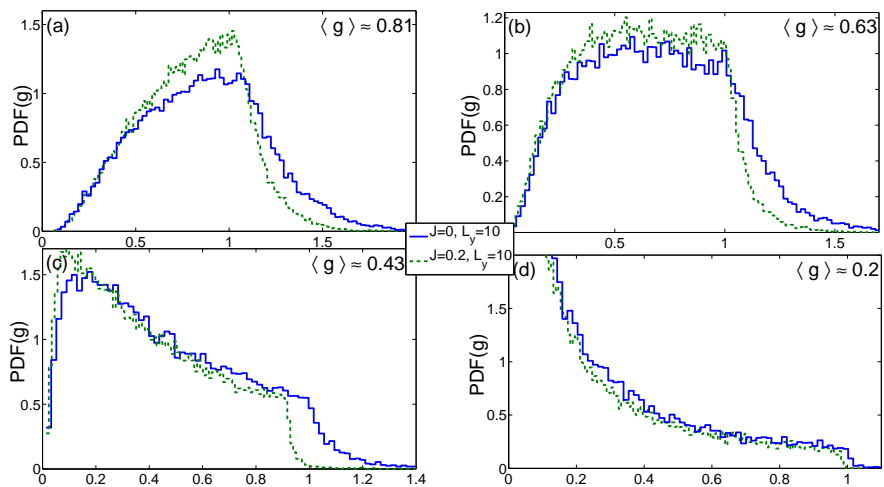}}
\caption{\label{fig:non_analytic}Comparison of Probability density functions (PDF) of conductance for $J=0$ (plain curves) and $J=0.2$ (dashed curves). Plots are performed for different values of average conductance. (a):  $\langle g \rangle(J=0)=0.84$ and $\langle g \rangle(J=0.2)=0.79$. (b):  $\langle g \rangle(J=0)=0.67$ and $\langle g \rangle(J=0.2)=0.62$. (c):  $\langle g \rangle(J=0)=0.45$ and $\langle g \rangle(J=0.2)=0.42$. (d):  $\langle g \rangle(J=0)=0.21$ and $\langle g \rangle(J=0.2)=0.18$.
}
\end{figure}
 \begin{figure}[ht!]
\centerline{\includegraphics[width=13cm]{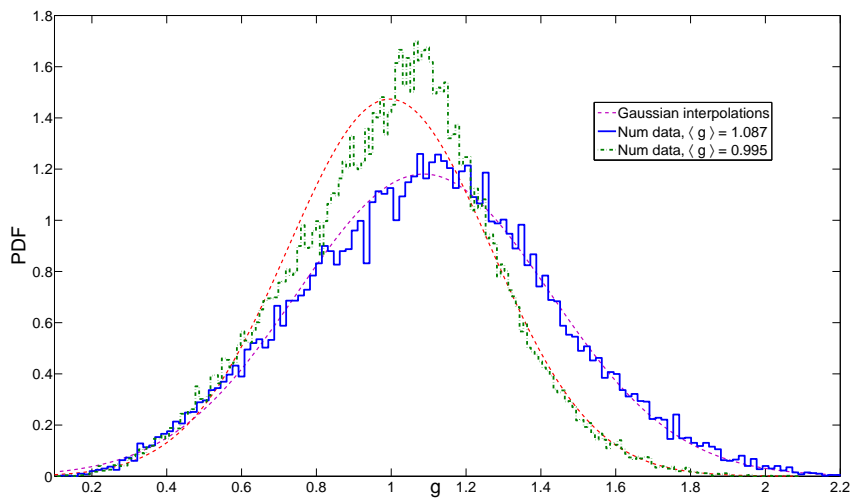}}
\caption{\label{fig:non_analytic_g1}PDF of conductance for $\langle g \rangle<1 (J=0.2)$ and $\langle g \rangle>1 (J=0)$ and Gaussian interpolations. $L_{y}=10$. 
}
\end{figure}
In order to study the dependance of this non-analyticity on the universality class, we have plotted on 
figure \ref{fig:non_analytic} the distribution $P(g)$ for similar values of $\langle g \rangle$
but different magnetic strengths  $J=0$ (GOE) and $J=0.2$ (GUE). 
 The shapes of these distributions are highly similar if $\langle g \rangle \ll 1$, 
 showing that distributions for $J=0$ and $J\neq 0$ reaches the same Log-normal distribution at large system sizes, in agreement with the super-universality scenario\cite{Qiao:2009}.
 In the intermediate regime ($\langle g \rangle\approx 1$), shapes are symmetry dependent. Moreover
  we find that the non-analyticity appears 
 for different values of conductance (close to $1$) and the rate of the exponential decay 
 \cite{Muttalib:1999} in the metallic regime seems to differ from one class to the other 
 (see for instance curves (a) or (b)). Finally, the plot on 
 figure \ref{fig:non_analytic_g1} represents the distribution of conductance of mean value just above and below the 
 threshold $\langle g \rangle =1$. Plain lines represent gaussian interpolations with a 
 mean and a variance given by the first and the second cumulant of each numerical conductance distribution. 
 For $\langle g \rangle>1$, the gaussian interpolation approximates very well the full 
 distribution while as soon as $\langle g \rangle<1$, the gaussian approximation only applies in the tail $g\geq 1$ of the distribution of conductance\cite{Muttalib:2003}. 
 %
The shape of the distribution for $g<1$ (Figure \ref{fig:non_analytic}) agrees qualitatively with the numerical results of \cite{Gopar:2002} (see in particular their fig. 4), 
Unfortunately, a more accurate comparison proves to be difficult due to the lack of analytical description of the distribution.
%
 To quantize further these results on the whole distribution of $\log g$ we now turn to a quantitative study of the second and third cumulant of this distribution.
 
 \subsection{Study of cumulants}
 
\begin{figure}[!t]
\centerline{
\includegraphics[width=13cm]{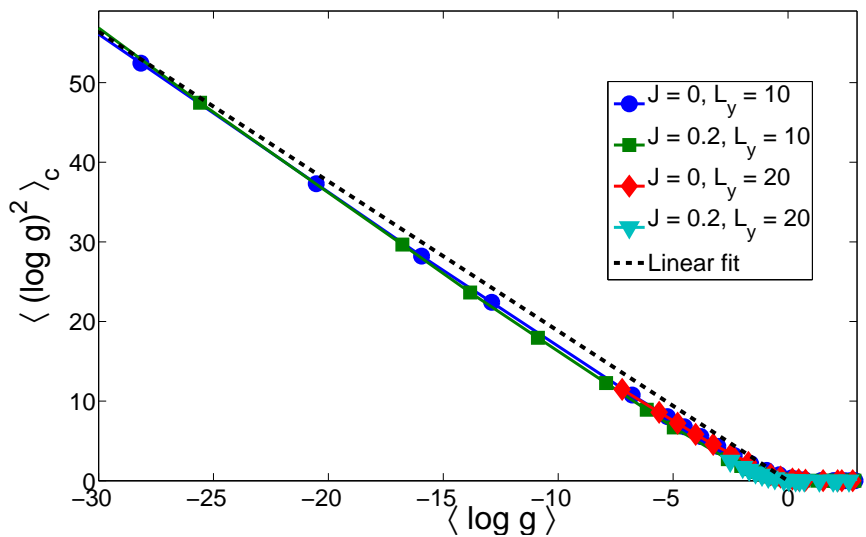}}
\caption{\label{fig:logGquad_logGmoy}
Plot of the variance of $\log g$ as a function of the mean for the orthogonal ($J=0$) and unitary ($J=0.2$) case, for $L_y=10$ and $20$. The slope of the linear fit is equal to $-1.88$. 
This plot shows also super-universality as the behavior of the second cumulant does not depend neither on geometry of the wire nor on the universality class. 
}
\end{figure}

 This conductance distribution converges to the Log-normal only deeply in the insulating regime, the convergence 
 being very slow (much slower than in the metallic regime). This qualitative result is 
 confirmed by the study of moments:  in the insulating regime the second cumulant is expected 
 to follow\cite{Beenakker:1991,Pichard:1991}:
\begin{equation}
\langle\left(\log g - \langle\log g\rangle\right)^2\rangle = \langle (\log g)^2\rangle_c = -2\langle\log g\rangle,
\label{eq:pasdaccord}
\end{equation}
Our numerical results are in agreement with this scaling with 
however very slow convergence towards this law:  corrections are measurable even for the largest system size where 
the system is deeply in the localized state, as shown on fig.~\ref{fig:logGquad_logGmoy}. 
More precisely, we find that for the deep insulating regime 
$\langle (\log g)^2\rangle_c = -1.88 \langle\log g \rangle $ slope $-1.88$, with a slight discrepancy with 
(\ref{eq:pasdaccord}).  
This plot also shows that in the deep insulating regime the behavior of the second cumulant as a function of the first one 
does not depend on the value of magnetic disorder:  both curves for $J=0$ and $J=0.2$ follow the same law. 
This is in agreement with our previous result on statistical distributions:  there is a super-universal behavior in 
the deep insulating regime.

\begin{figure}[!t]
\centerline{
\includegraphics[width=13cm]{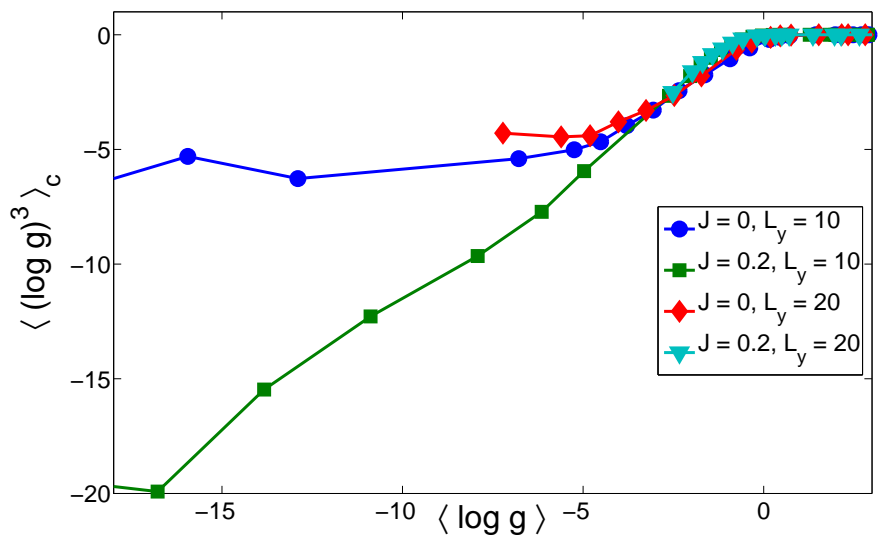}}
\caption{\label{fig:logGcube_logGmoy}
Plot of the skewness of $\log g$ as a function of the mean for the orthogonal ($J=0$) and unitary ($J=0.2$) case, for $L_y=10$ and $20$.
}
\end{figure}

Finally  we have studied the third cumulant of $\log g$ scaled in figure \ref{fig:logGcube_logGmoy} as a function of the first cumulant. 
The linear behavior for each value of magnetic disorder in the deep insulating regime ($\langle \log g \rangle<-4$) 
is in agreement with the single parameter scaling. We find that 
contrary to the second cumulant the coefficient of proportionality between the skewness and the average depends on 
the symmetry of disorder, which denotes a lack of super-universality concerning this cumulant. 
For instance dots and diamonds (which correspond to the case $J=0$) have the same behavior, 
as opposed to the case $J=0.2$ (squares or triangles). A systematic study of this point with even larger system sizes and other numerical methods would be of high interest but is definitely beyond the scope of the present paper. 

The study of cumulants of the distribution of $\log g$ confirm the single parameter scaling of the distribution, with 
a slight discrepancy concerning the value of the coefficient of proportionality between second and first cumulant. 
Moreover, super-universality has been highlighted concerning the second cumulant but is lacking concerning the 
third one.

\subsection{Comparison with exact results in the cross-over regime}

\begin{figure}[!t]
\centerline{
\includegraphics[width=13cm]{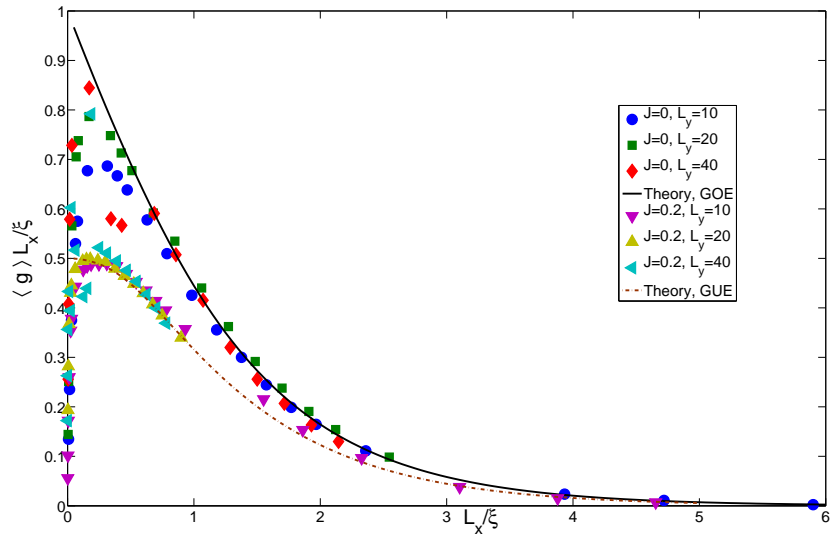}}
\caption{\label{fig:gmoyenGOE_GUE}
Comparison of the numerical average conductance $\langle g \rangle$ as a function of the reduced length $L_{x}/\xi$ with the exact 
expressions of ref.~\cite{Mirlin:1994} for the  orthogonal ($J=0$) and unitary ($J=0.2$) case, for various transverse sizes 
$L_y$.
}
\end{figure}

\begin{figure}[!t]
\centerline{
\includegraphics[width=13cm]{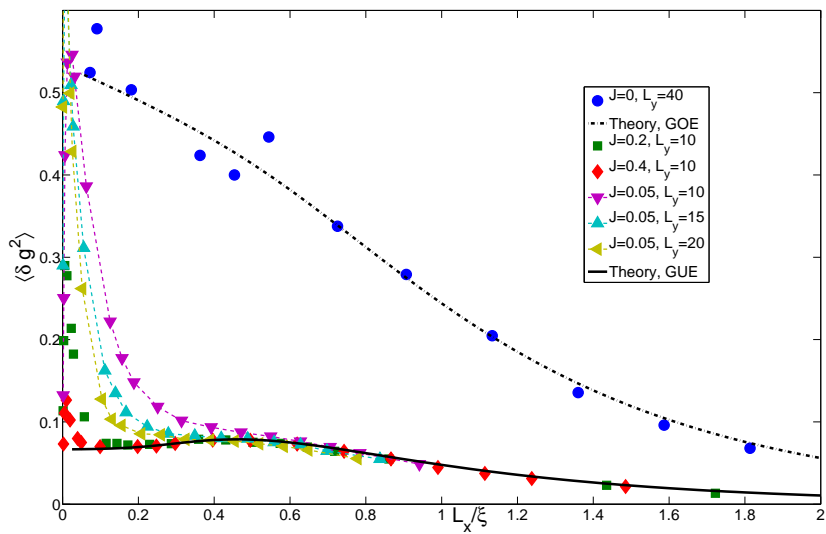}}
\caption{\label{fig:varianceGOE_GUE}
Comparison of the numerical average conductance fluctuations $\langle \delta g^{2} \rangle$ 
as a function of the reduced length $L_{x}/\xi$ with the exact 
expressions of ref.~\cite{Mirlin:1994} for the  orthogonal ($J=0$) and unitary ($J=0.4$) case, for various transverse sizes 
$L_y$.

}
\end{figure}

For localization in wires connected to ideal contacts, 
exact formula for the average conductance \cite{Zirnbauer:1992} and conductance fluctuations 
\cite{Mirlin:1994} have been derived 
for the two universal orthogonal and unitary classes. These formula are of particular interest in the intermediate regime between the metallic ($L_{x}\ll \xi$) and the deeply localized ($L_{x}\gg \xi$) regime. 
 We have numerically evaluated the formula (3.105) and (3.106) of ref.~\cite{Mirlin:1994} and we compare them with our numerical data  in Fig.~\ref{fig:gmoyenGOE_GUE} and Fig.~\ref{fig:varianceGOE_GUE}. 
 We find an excellent agreement between the $J=0$ data and the orthogonal exact formulae on one hand, 
 and the $J\neq 0$ 
 data and the unitary formulae on the other hand. This comparison naturally breaks down for small sizes    where the quasi-1d assumption for diffusion  breaks down and a non universal regime takes place. 
As shown on Fig.~\ref{fig:varianceGOE_GUE}, the exact unitary behavior is recovered for $J=0.05$ beyond the cross-over length 
$L_{m}$ (see next section). Whether the scale dependance of $\langle g \rangle$ and 
$\langle \delta g^{2} \rangle$ in the cross-over regime ($J$ small such that $L_{m} \simeq \xi$) 
 is amenable to an exact formula along the lines of ref.~\cite{Mirlin:1994} is an open question of high interest. 

\section{Universal Metallic Regime}

We now focus on the universal metallic regime described by weak localization. By definition weak localization corresponds to metallic diffusion, expected for lengths of wire $l_{e}\ll L_{x}\ll \xi$. 
For this regime to be reached, we thus need to increase  $\xi$ through an increase of the 
 number of transverse modes $L_{y}$ with all other parameters fixed (see eq.~(\ref{equ:xi_gen})). 
 Moreover, for a fixed 
geometry, this regime will be easier to reach with magnetic impurities than without. 
 As we saw on fig.~\ref{fig:non_analytic}, the shape of the PDF of conductance 
is a truncated gaussian in this regime. In the following we study its three first cumulants quantitatively, starting with the variance.

\subsection{Conductance Fluctuations and Universal Crossover}

\begin{figure}[ht]
 \centerline{
\includegraphics[width=13cm]{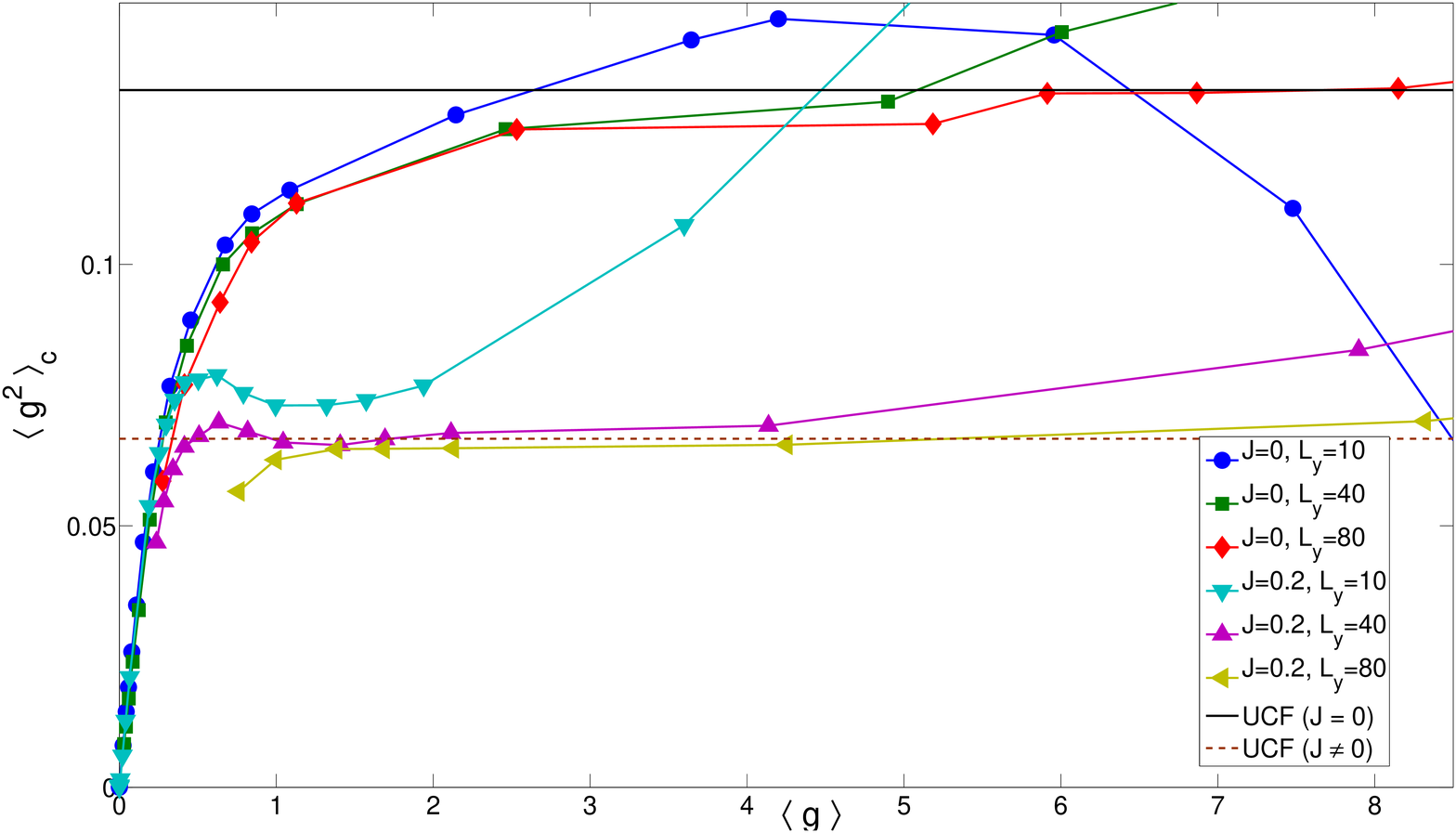}
}
\caption{\label{fig:Gquad_Gmoy}
Second cumulant of $g$ as a function of first cumulant of $g$ showing the universal behavior in the metallic regime. Different curves correspond to different transverse lengths ($L_y=10, 40, 80$) or magnetic disorder ($J=0, 0.2$).
}
\end{figure}

\begin{figure}[ht]
\centerline{
\includegraphics[width=13cm]{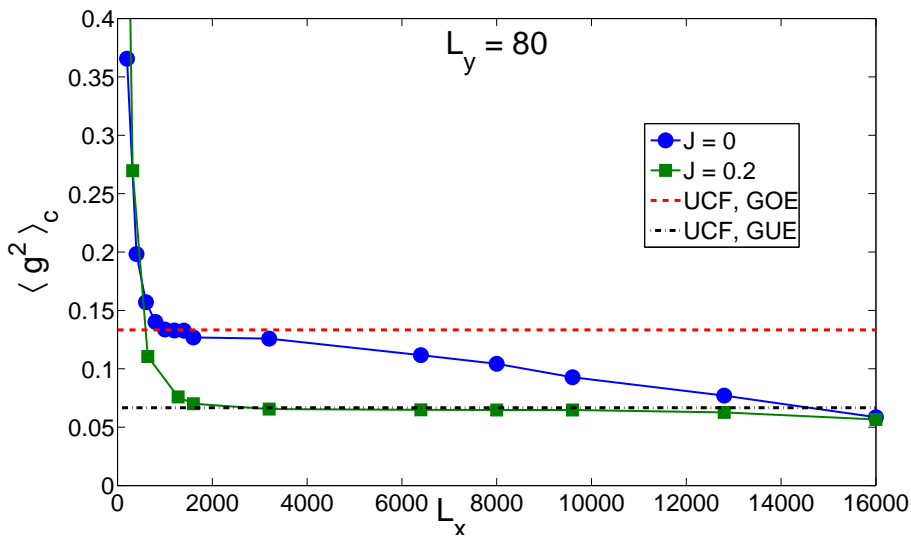}
}
\caption{\label{fig:GquadLy80_Lx}
Second cumulant of $g$ as a function of longitudinal length, for $L_y=80$. Dots represent data for $J=0$ and squares for $J=0.2$. 
The value of UCF is shown in each symmetry class (with or without magnetic disorder). UCF regime is reached in both cases.
}
\end{figure}

\begin{figure}[ht]
\centerline{
\includegraphics[width=13cm]{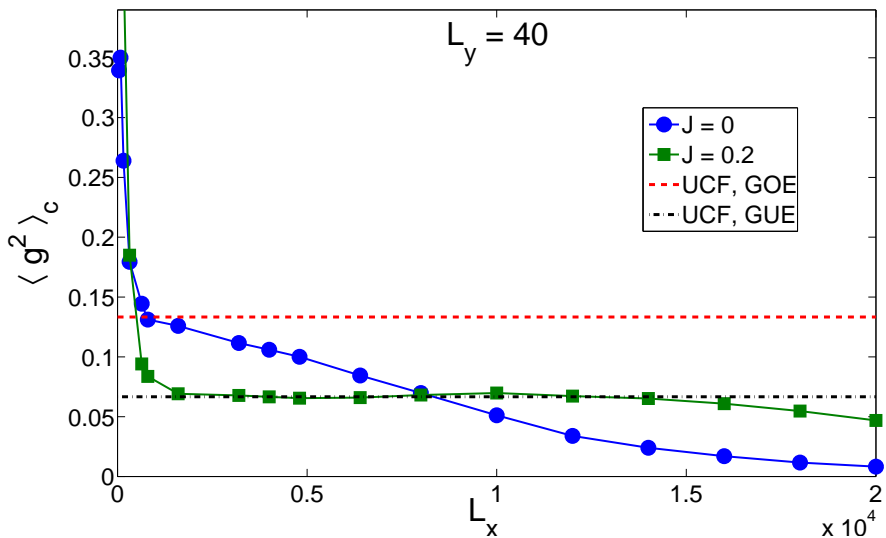}
}
\caption{\label{fig:GquadLy40_Lx}
Second cumulant of $g$ as a function of longitudinal length, for $L_y=40$. Dots represent data for $J=0$ and squares for $J=0.2$. 
The value of UCF is shown in each symmetry class (with or without magnetic disorder. UCF regime is reached in the case of magnetic impurities but not for scalar impurities.
}
\end{figure}

In the weak localization regime, the conductance fluctuations $\langle g^2 \rangle_c$ are expected to be independent on the size of the system, and only depend on the universal localization class of the model. 
The figure \ref{fig:Gquad_Gmoy} shows that for a suitable value of transverse length, 
the system reaches the expect plateau in conductance fluctuations. The value of the plateau identifies with the expected values ($1/15$ and $4/15$ for GUE and GOE respectively) with a high accuracy.  
 However, the presence of this plateau depends strongly on the value of transverse length $L_y$ and on 
the magnetic disorder. 
Figures \ref{fig:GquadLy80_Lx} and \ref{fig:GquadLy40_Lx} illustrates this point with further details:  
these plots show the conductance fluctuations as a function of longitudinal size 
for $J=0$ and $J=0.2$ and for two values of transverse length $L_y$. On the first plot, for both values of magnetic 
disorder the universal plateau arises, whereas it appears only for $J=0.2$ if $L_y=40$.

\begin{figure}[ht]
\centerline{
\includegraphics[width=13cm]{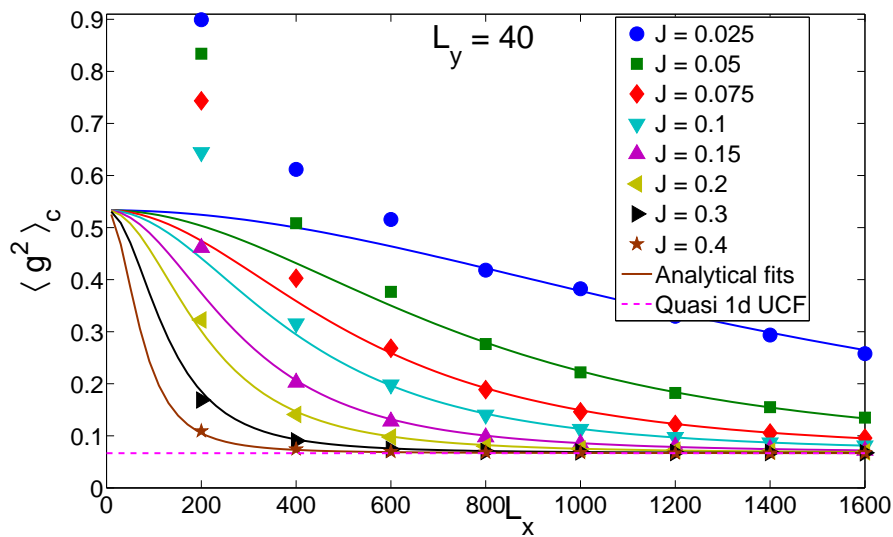}
}
\caption{\label{fig:Gquad_Lx}
Variance of $g$ as a function of longitudinal size. UCF are shown. Different curves correspond to different 
values of magnetic disorder $J$ ($J=0.025\to 0.4$). Transverse length $L_y=40$. The only free parameter in analytical fits is the magnetic length $L_m$.
}
\end{figure}

\begin{figure}[ht]
\centerline{
\includegraphics[width=13cm]{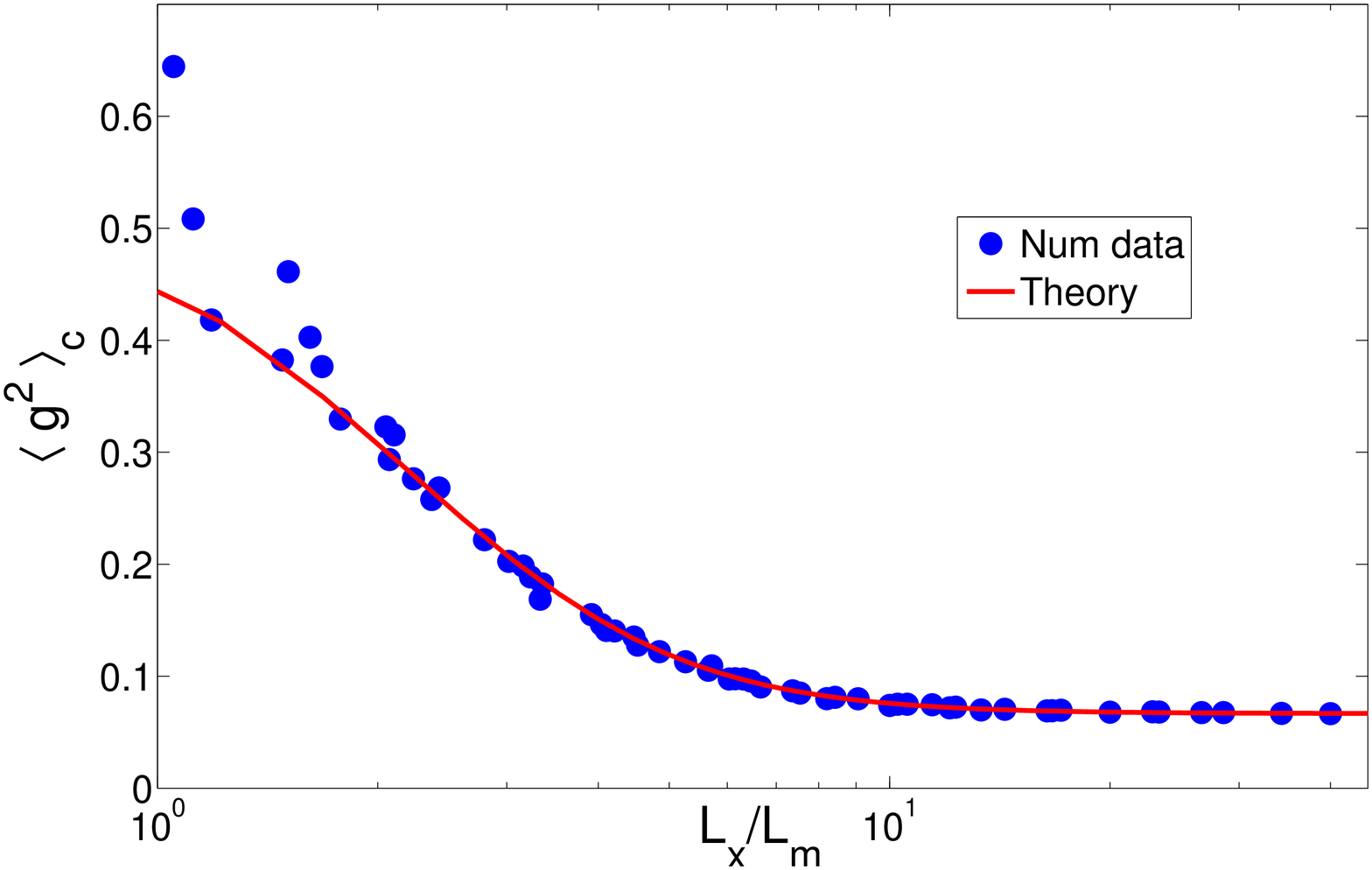}
}
\caption{\label{fig:Gquad_LxoverLm}
Variance of $g$ as a function of $L_x/L_m$. Transverse length $L_y=40$.
}
\end{figure}

The evolution of this variance of the PDF of $g$ depends on the longitudinal length through  
\cite{Akkermans:2007}: 
\begin{eqnarray}\label{equ:theo_ucf}
\langle\delta g^2\rangle = \langle g^2 \rangle_c &=&  \frac{1}{4}F\left(0\right) + \frac{3}{4}F\left(x\sqrt{\frac{4}{3}}\right)  
 + \frac{1}{4}F\left(x\sqrt{2}\right) + \frac{1}{4}F\left(x\sqrt{\frac{2}{3}}\right),
\end{eqnarray}
where $x = L_x/L_m$ and the scaling function $F(x)$ depends only on dimension\cite{Akkermans:2007,Paulin:2011b}. 
The universality occurs in this equation in the two limit 
$x=0$ corresponding to $J=0$ or  $x\gg 1$ where $F(x)\to 0$. In both cases the variance becomes  
geometry independent. 
 Moreover this expression shows that the whole crossover between the two classes is described by a universal crossover function parametrized solely by the length $L_{m}$, called the magnetic dephasing length scale \cite{Akkermans:2007}. 
 In the figure \ref{fig:Gquad_Lx}, these conductance fluctuations are plotted as a function of longitudinal length $L_{x}$ for different values of $J$. A single parameter fit  by (\ref{equ:theo_ucf}) provides the determination of the magnetic dephasing length $L_m$ 
as a function of the magnetic disorder $J$. The Universal behavior is also highlighted for strong enough magnetic disorder. 
The determination of this scattering length allows a precise study of average conductance, 
and in particular the study of the classical part as described in the next sub section below. Fig.~\ref{fig:Gquad_LxoverLm} shows the scaling form of these 
fluctuations (as a function of $L_{x}/L_{m}(J)$) in excellent agreement with the theory (\ref{equ:theo_ucf}). 
 Moreover,  for long wires (and large values of $J$) conductance fluctuations are no longer $L_x$ dependent and equal to $1/15$, in units of $G_0^2$. 
 This is the so-called Universal Conductance Fluctuations (UCF) regime which is precisely identified numerically in the present work.
The plot of figure \ref{fig:Gquad_Gmoy} confirms analytical results from \cite{Froufe:2002} 
both qualitatively in the shape of the curves and quantitatively in the values of fluctuations in both universality classes. 
In our study, values of UCF are reached with a maximal error of $1 \%$ for GOE and $3 \%$ for GUE with respect to 
the analytical value of the UCF in the regime independent  of $\langle g \rangle$ ({\it i.e} with much higher precision 
 than {\it e.g} \cite{Cieplak:1991} and \cite{Markos:2002}). 

Let us finally comment the work of Z. Qiao {\it et al.} \cite{Qiao:2009} who have performed a similar numerical Landuer study of 1D transport in various universality classes, focusing mostly on the metallic regime. While both our study agree on the UCF (although we have a higher accuracy for $\beta=1$), we did not find evidence for  a second universal conductance plateau. This result would definitely deserves further study. 

\subsection{The average conductance}

\begin{figure}[!t]
\centerline{
\includegraphics[width=13cm]{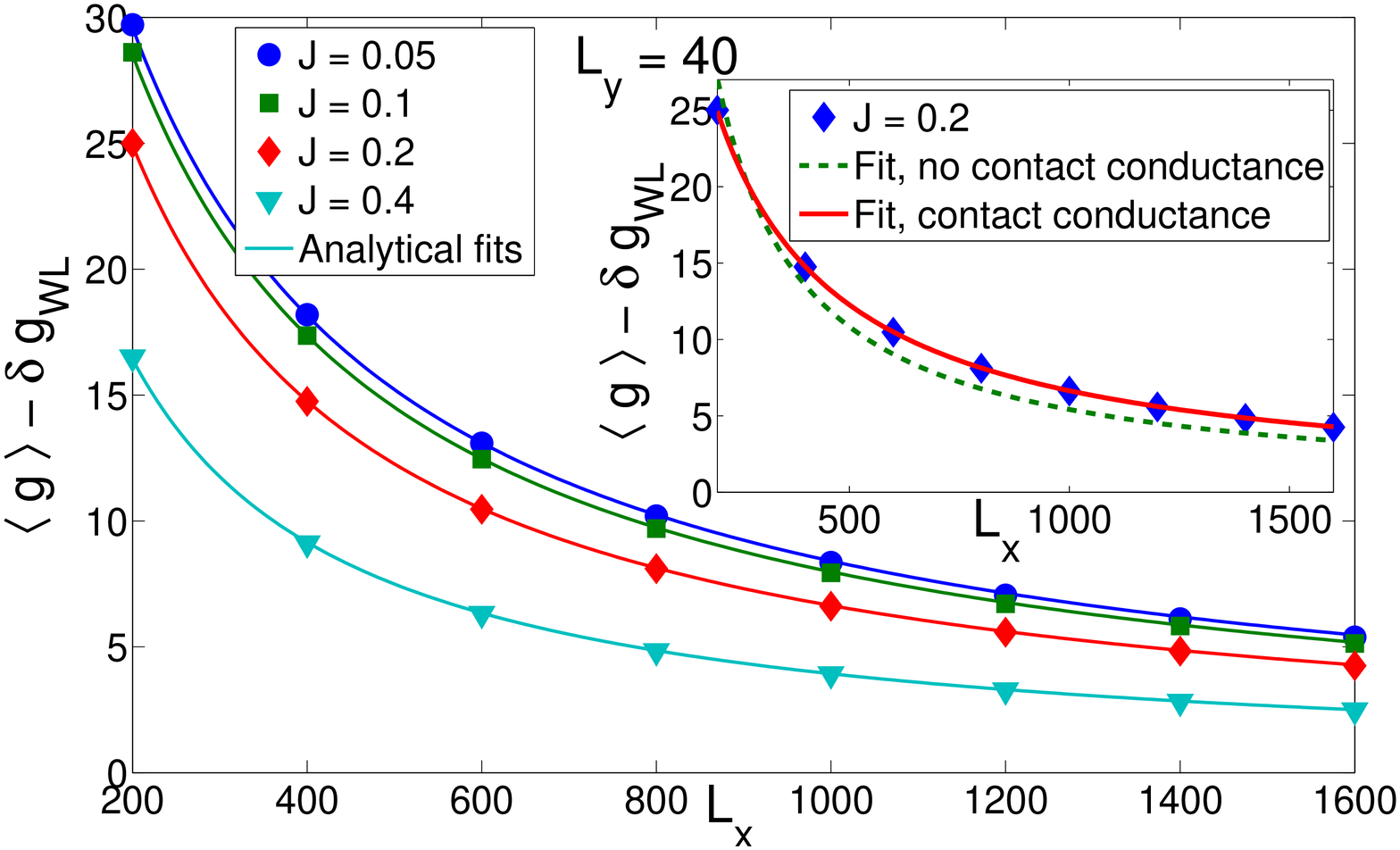}}
\caption{\label{fig:classical_conductance}
Evolution of classical conductance $g_{class}=\langle g \rangle-\delta g$ with the longitudinal length for $L_y=40$ and $J=0.05, 0.1, 0.2, 0.4$. 
Numerical data are well fitted using the conductivity as the only free parameter. different data correspond 
to different values of magnetic disorder. In inset is shown the set of numerical data for $J=0.2$. The dotted line 
is the analytical fit without taking into account the contact conductance. Plain line is the same analytical fit as in the main plot.
}
\end{figure}

\begin{figure}[!t]
\centerline{
\includegraphics[width=13cm]{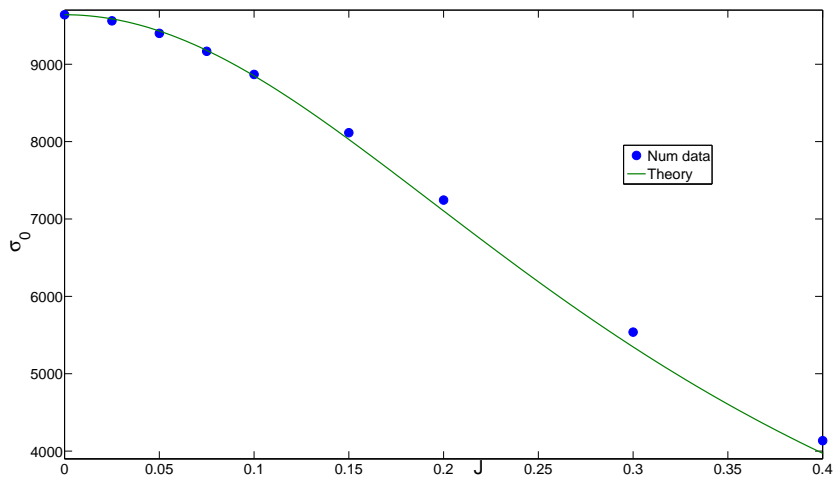}}
\caption{\label{fig:sigma0}
Evolution of conductivity $\sigma_0$ with magnetic disorder $J$ for $L_y=40$. Dots are numerical data. Plain line is the theory given by 
Einstein relation and Matthiesen rule for the conductivity. Agreement is good especially at low magnetic disorder.
}
\end{figure}

The main contribution to the average conductance is  of classical origin. However a weak localization correction must be added when 
the quantum behavior of electrons is taken into account \cite{Akkermans:2007}. This quantum part manifests itself in the magneto-conductance behavior of long wires (larger than the the phase coherence length)  
 where a weak magnetic field is sufficient to destroy this quantum correction by dephasing the various diffusing path with respect to each other\cite{Akkermans:2007}. 
 We can write this average conductance as (without any magnetic field):  
\begin{equation}
\langle g \rangle(J,L_x,L_m) = g_{class}(J,L_x) + \delta g_{WL}(L_x,L_m),
\label{equ:averageG}
\end{equation}
where $g_{class}$ is the classical part of the conductance and $\delta g_{WL}$ is the weak localization correction. 
For a quasi one dimensional system the quantum correction reads the simple form\cite{Akkermans:2007}
\begin{equation}
\delta g_{WL} = \sum_{n=1}^{\infty}\left(\frac{-1/\pi^2}{n^2+2\left(\frac{L_x}{L_m}\right)^2}-\frac{3/\pi^2}{n^2+\frac{2}{3}\left(\frac{L_x}{L_m}\right)^2}\right).
\end{equation}
The knowledge of the magnetic length $L_m$ we gained in the previous study of conductance fluctuations can now be used to completely characterize this weak localization contribution to the conductance. By subtracting the corresponding contribution to the average conductance, we obtain the classical conductance, plotted in figure \ref{fig:classical_conductance} 
as a function of longitudinal length. 
Taking into account the contact resistance\cite{Engquist:1981} in the two terminal setup, the expected expression for this classical conductivity reads 
\begin{equation}
g_{class}(J,L_x) = \frac{1}{\frac{1}{L_y}+\frac{L_x}{\sigma_0(J)}}.
\label{equ:class_cond}
\end{equation}
Figure \ref{fig:classical_conductance} shows this expression plotted for different values of magnetic disorder. 
The corresponding fitting parameter $\sigma_{0}$ is plotted  as a function of $J$ in figure \ref{fig:sigma0}.
To compare these results with theory, consider the Einstein relation which links the  (Einstein) conductivity to the diffusion constant:
\begin{equation}
\sigma_0 = s e^2\rho_0(\varepsilon_F)D,
\end{equation}
where $s$ is the spin degeneracy and $\rho_0(\varepsilon_F)$ the electronic density of states at the Fermi level. 
By definition, the diffusion coefficient reads, for non magnetic impurities:
\begin{equation}
D = v_F^2\tau_e,
\end{equation}
with $v_F$ the Fermi velocity and $\tau_e$ the elastic scattering time. It can be related to the scalar disorder by:
\begin{equation}
\tau_e = \frac{1}{2\pi\rho_0n_iv_0^2}
\end{equation}
where $n_i$ is the impurity density and $v_0^2 = W^2/12$. For more than one diffusive process, 
it is compulsory to use the Matthiesen rule that modifies scattering time $\tau_e$ in the following way:
\begin{equation}
\frac{1}{\tau_e}\to\frac{1}{\tau_e}+\frac{1}{\tau_m},
\end{equation}
where $\tau_m=L_m^2/D$ and is related to the magnetic disorder:
\begin{equation}
\tau_m = \frac{1}{2\pi\rho_0n_iJ^2\langle S^2\rangle}.
\end{equation}
Using this allows one to get the $J$ dependance of the Einstein conductivity:
\begin{equation}
\sigma_0(J) = \frac{\sigma_0(J=0)}{1 + \frac{3}{W^2}J^2}.
\end{equation}
 In figure \ref{fig:sigma0} we compare this expression 
with numerical evaluation of the conductivity. The good agreement between both curves provides an additional check of the correct determination of the magnetic dephasing length $L_{m}$. 
Including the magnetic disorder dependance of the diffusion coefficient (through  Matthiesen rule), we 
obtain a perturbative expression to second order in $J$ for this magnetic dephasing length:
\begin{equation}
\label{eq:LmPerturb}
L_m(J) = \sqrt{D(J)\tau_m(J)}\propto \frac{1}{J\sqrt{\frac{W^2}{12}+\frac{J^2}{4}}}.
\end{equation}
\begin{figure}[!t]
\centerline{
\includegraphics[width=13cm]{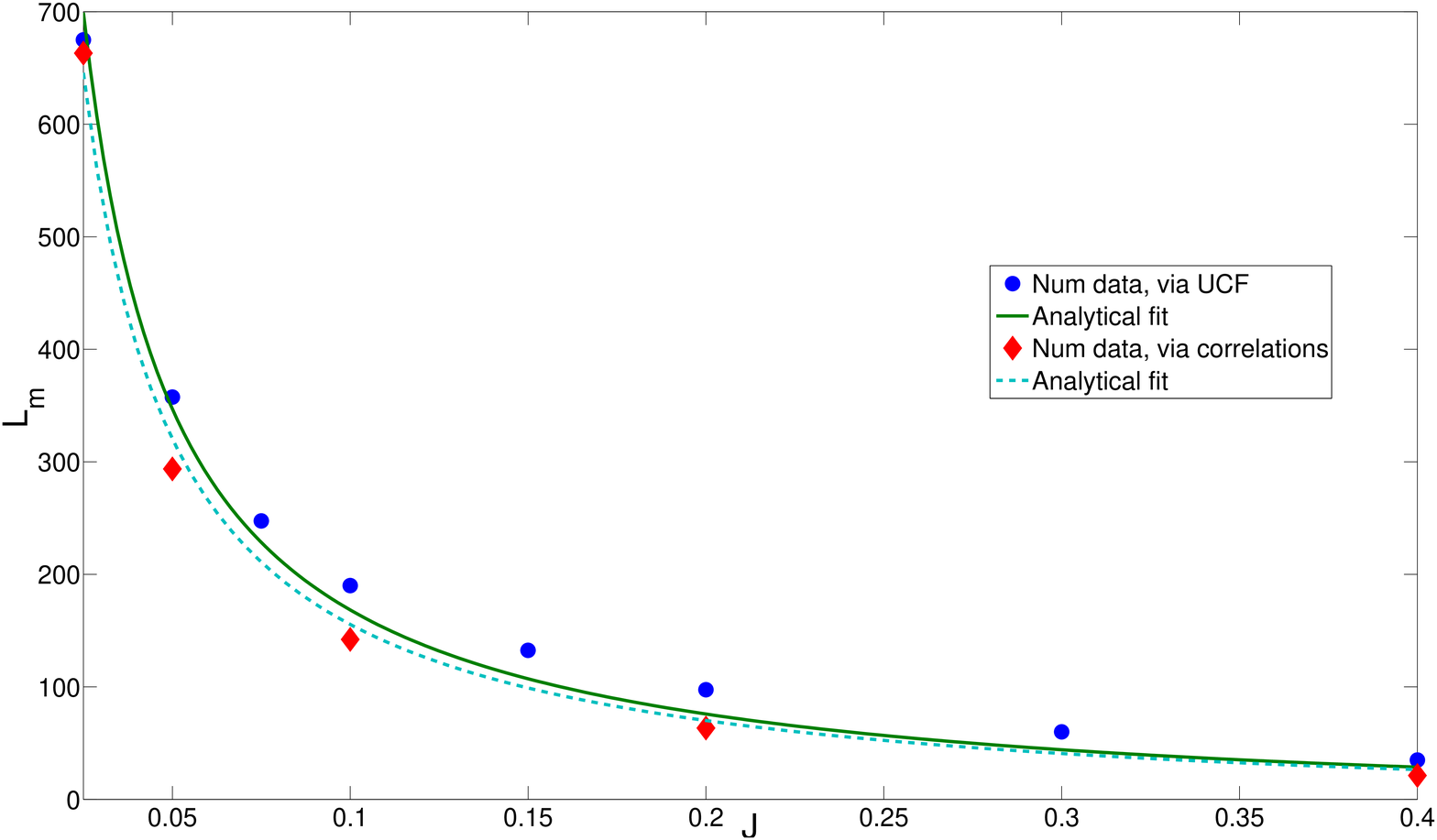}}
\caption{\label{fig:Lm_J}
Evolution of magnetic dephasing length $L_m$ with magnetic disorder $J$ for $L_y=40$. Dots are numerical data extracted from 
the study of UCF and diamonds are numerical data extracted from the study of conductance correlations for
 two different spin configurations. Plain and dotted lines are analytical fits from perturbation theory at second 
 order in $J$. Error bars are smaller than dots and diamonds sizes.
}
\end{figure}
On figure \ref{fig:Lm_J} we have plotted the numerical evaluation of the magnetic length as a
function of magnetic disorder and the corresponding fit with eq.(\ref{eq:LmPerturb}). 
We
notice that it is also possible to obtain the magnetic length via the
study of correlations of conductance, \textit{i.e} via the study of
$\langle g(V,\{\vec{S}_i^{(1)}\}_i)g(V,\{\vec{S}_i^{(2)}\}_i)\rangle_c$, which goes beyond the scope of the present paper\cite{Paulin:2011b}. 


\subsection{The third cumulant}

Finally we consider the third cumulant of the distribution of conductance. 
According to the analytical study of \cite{Froufe:2002}, this cumulant decays to zero in a universal way
as $\langle g\rangle$ increases. Here in figure \ref{fig:Gcube} we find a dependance of this decrease 
on the symmetry class:  for GOE $\langle g^3 \rangle_c$ goes to zero in a monotonous way whereas it 
decreases, changes its sign and then goes to zero in GUE case. For $\langle g \rangle>4$ numerical errors 
are dominant, then this part of the curve is irrelevant. Moreover, for GUE this decrease seems to be universal 
whereas it depends on the transverse length for GOE.
  \begin{figure}[ht]
\includegraphics[width=13cm]{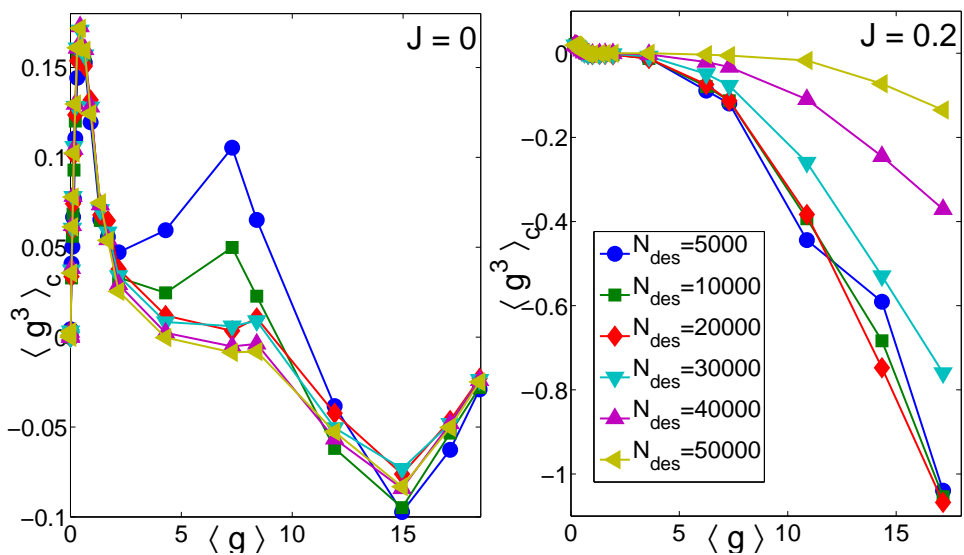}
\caption{\label{fig:convergence}
Plot of $\langle g^3 \rangle_c$ as a function of $\langle g \rangle$ in the metallic regime, averages are performed with various a number of configurations $N_{des}$. Convergence curves are shown for $L_y=10$ and $J=0$ or $J=0.2$.}
\end{figure}
On figure \ref{fig:convergence}, is represented the convergence of the skewness when increasing the number of configurations used to perform averages $N_{des}$ for both GOE and GUE. Plots show a good enough convergence of averages to conclude that the third cumulant of conductance is not zero for all values of $\langle g \rangle$. Notice that the maximal number of averages is 50000.
\begin{figure}[ht]
\includegraphics[width=13cm]{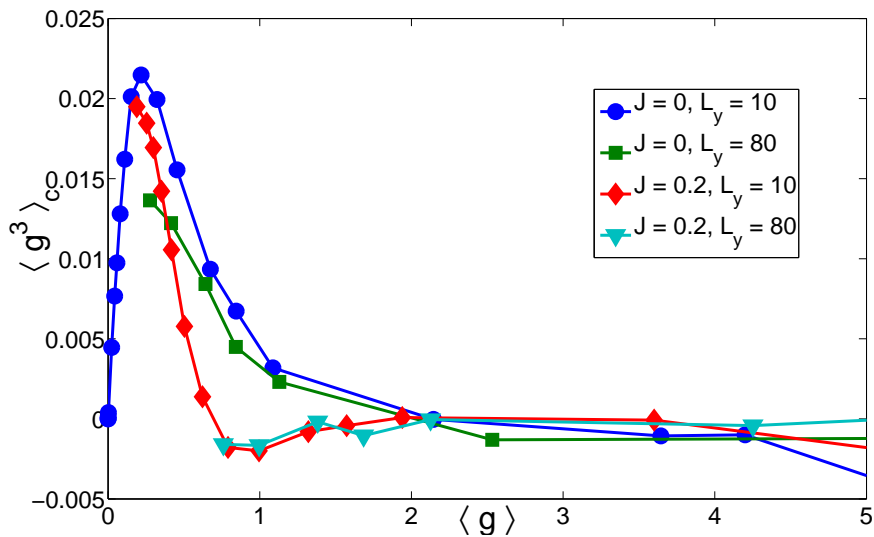}
\caption{\label{fig:Gcube}
Plot of $\langle g^3 \rangle_c$ as a function of $\langle g \rangle$ in the metallic regime for $L_y=10, 80$ and $J=0, 0.2$.}
\end{figure}
Moreover this fast vanishing of the third cumulant confirms the faster convergence of the whole 
distribution towards the gaussian, compared to what happens in the insulating regime. 
Based on our numerical results, we cannot confirm nor refute the expected 
law $\langle g^3 \rangle_c \propto 1/\langle g \rangle^n$, with $n=2$ in GOE and $n=3$ in GUE \cite{Macedo:1994,vanRossum:1997}.


\section{Conclusion}

To conclude we have conducted extensive numerical studies of electronic transport in the presence of 
random frozen magnetic moments. Comparing and extending previous analytical and numerical studies, we 
have identified the insulating and metallic regimes described by the universality classes GOE and GUE. We have 
paid special attention to the dependance on this symmetry of cumulants of the distribution of conductance in 
both metallic and insulating universal regimes. In particular, we have identified with high accuracy the 
domain of universal conductance fluctuations, and determined its extension in the present model. We have also determined precisely the so-called magnetic length $L_m$ which represents the elastic scattering length of the spin on magnetic impurities. This length is of primary importance in experiments as it controls the crossover between universality classes. 
This work paves the way for further studies of transport in metals with frozen magnetic impurities as we have clearly identified the range of the parameters to access the experimentally relevant metallic diffusive regime. One possible extension consists in considering evolution of the statistics of conductance as  the magnetic disorder is varied, \textit{e.g.} by rotating or flipping the spins of impurities. Comparing the conductance obtained in both spin configurations mimics the experimental measurement of the conductance of a low temperature canonical spin glass after two successive quenches \cite{Carpentier:2008, Paulin:2011b}, 
without the necessary restrictions of analytical approaches\cite{Paulin:2011b}. 
Experimentally, this approach could give access to fundamental properties of a spin glass, that have never been measured.


We thank X. Waintal for useful discussions.
This work was supported by the  ANR grants QuSpins and Mesoglass. All numerical calculations 
were performed on the computing facilities of the ENS-Lyon calculation center (PSMN).

\bibliographystyle{unsrt}
\bibliography{Meso,SpinGlass,MesoGlass}

\end{document}